\documentclass{article}

\usepackage[figuresright]{rotating}

\usepackage{longtable}
\usepackage{graphicx}
\usepackage{booktabs}
\usepackage{amsfonts, amsmath,amssymb,amsthm}
\usepackage{subcaption}
\usepackage{relsize}
\usepackage[hyperindex=true,colorlinks=true,urlcolor=black,citecolor=black,linkcolor=black]{hyperref}
\usepackage{multirow}
\usepackage{array}
\newcolumntype{M}[1]{>{\centering\arraybackslash}m{#1}}
\newcolumntype{P}[1]{>{\raggedright\arraybackslash}p{#1}}
\usepackage{authblk}
\usepackage{enumerate}
\usepackage[left=1.in,right=1.0in,top=1.5in,bottom=1.0in,includefoot,marginparwidth=1in, marginparsep=0.1in]{geometry} 

\newcommand\revision[1]{#1}

\newcommand{\Intersection}{\mathcal{I}}
\newcommand{\be}{\begin{equation}}
\newcommand{\ee}{\end{equation}}

\newcommand{\fracd}[2]{{\displaystyle \frac{#1}{#2}}}

\newcommand{\Rout}{R_{\textnormal{out}}}
\newcommand{\ConvOp}{{\bf C}_{\convWeight}}
\newcommand{\nDensityBulk}{\nDensity_{\textnormal{b}}}
\newcommand{\nDensityContact}{\nDensity_{\textnormal{C}}}

\newcommand{\HalfSpace}{H}
\newcommand{\Disc}{\mathcal{D}_1}
\newcommand{\InfAnn}{\mathcal{D}_2}
\newcommand{\FinAnn}{\mathcal{D}_3}
\newcommand{\Ball}{\mathcal{S}_1}
\newcommand{\Sphere}{\mathcal{S}_2}
\newcommand{\AnnSeg}{\mathcal{P}_1}
\newcommand{\HalfStripMinusDisc}{\mathcal{P}_2}
\newcommand{\SlitMinusDisc}{\mathcal{P}_3}
\newcommand{\WedgeCutSide}{\mathcal{P}_4}
\newcommand{\WedgeCut}{\mathcal{P}_5}
\newcommand{\FiniteWedge}{\mathcal{P}_6}
\newcommand{\Slit}{\text{slit}}

\newcommand{\weight}{\omega}
\newcommand{\PhysSpace}{\mathcal{Y}}
\newcommand{\AlgMapWhole}{\mathcal{A}_1}
\newcommand{\AlgMapSemi}{\mathcal{A}_2}
\newcommand{\AlgMapSemiFinite}{\mathcal{A}_{2,\textnormal{F}}}

\newcommand{\CartMap}{\mathcal{C}}
\newcommand{\CartMapSph}{\CartMap_{\text{sph}}}

\newcommand{\eD}{\LJdepth_{\text{D}}}

\newcommand{\PhiTwoD}{\Phi_{\text{2D}}}

\newcommand{\PhiTwoDAttr}{\PhiTwoD^{\text{attr}}}

\newcommand{\UpsHS}{\Upsilon_{\text{HS}}}
\newcommand{\UpsHD}{\Upsilon_{\text{HD}}}
\newcommand{\HD}{{\text{HD}}}
\newcommand{\ytwomax}{y_{2,\text{max}}}
\newcommand{\merr}{m_{\text{err}}}

\newcommand{\relaxationPar}{\lambda_r}

\newcommand{\convWeight}{\chi}
\newcommand{\xE}{\hat x}

\newcommand{\refe}[1]{(\ref{#1})}

\newcommand\etal{\mbox{\textit{et al.}}}

\newcommand{\diff}[2]{\frac{\partial #1}{\partial #2}}

\newcommand{\klamm}[1]{\left( #1 \right)}
\newcommand{\klammCurl}[1]{\left\{ #1 \right\}}

\newcommand{\Grad}{{\boldsymbol \nabla}}

\newcommand{\dI}{\text{d}}
\newcommand{\ofR}{\klamm{\bf r}}

\newcommand{\defi}{=}

\newcommand{\nDensity}{n}

\newcommand{\LJdepth}{\varepsilon}
\newcommand{\LJWdepth}{\LJdepth_{\text{w}}}
\newcommand{\LJdiam}{\sigma}

\newcommand{\surfaceTension}{\gamma}

\newcommand{\pos}{{\bf r}}

\newcommand{\frictionCoeff}{\gamma}

\newcommand{\DisjoiningPressure}{\Pi}
\newcommand{\chemPotN}{\mu}

\newcommand{\nDensityV}{\nDensity_{\text{vap}}}
\newcommand{\nDensityL}{\nDensity_{\text{liq}}}

\newcommand{\surfaceTensionLV}{\surfaceTension_{\text{lv}}}
\newcommand{\surfaceTensionWV}{\surfaceTension_{\text{wv}}}
\newcommand{\surfaceTensionWL}{\surfaceTension_{\text{wl}}}

\newcommand{\GrandPotential}{\Omega}

\newcommand{\DL}[1]{\tilde{#1}}

\newcommand{\FE}{\mathcal{F}}
\newcommand{\FEexc}{\mathcal{F}_{\text{exc}}}
\newcommand{\FEideal}{\mathcal{F}_{\text{id}}}
\newcommand{\FEhs}{\mathcal{F}_{\text{HS}}}
\newcommand{\FEattr}{\mathcal{F}_{\text{attr}}}

\newcommand{\Vext}{V_{\text{ext}}}

\newcommand{\BHattr}{\phi_{\text{attr}}}

\newcommand{\depthLJ}{\varepsilon}
\newcommand{\depthLJW}{\depthLJ_{\text{w}}}

\newcommand{\phiLJ}{\phi_{\text{LJ}}}

\newcommand{\grad}[1]{{\boldsymbol \nabla}_{#1}}

\begin{document}


\title{Pseudospectral methods for density functional theory in bounded and unbounded domains}

\author[1]{Andreas Nold\footnote{Theory of Neural Dynamics, Max Planck Institute for Brain Research, 60438 Frankfurt am Main, Germany}}
\author[2]{Benjamin D. Goddard}
\author[1]{Peter Yatsyshin} 
\author[3]{Nikos Savva}
\author[1]{Serafim Kalliadasis\footnote{Corresponding author. Email address: \url{s.kalliadasis@imperial.ac.uk}}}

\affil[1]{Department of Chemical Engineering, Imperial College London, London, SW7 2AZ, UK}
\affil[2]{School of Mathematics and Maxwell Institute for Mathematical Sciences, The University of Edinburgh, Edinburgh, EH9 3FD, UK}
\affil[3]{School of Mathematics, Cardiff University, Cardiff CF24 4AG, United Kingdom}
\maketitle
\begin{abstract}
Classical Density Functional Theory (DFT) is a statistical-mechanical framework to analyze fluids, which accounts for nanoscale fluid inhomogeneities and non-local intermolecular interactions.
DFT can be applied to a wide range of interfacial phenomena, as well as problems in adsorption, colloidal science and phase transitions in fluids. Typical DFT equations are highly non-linear, stiff and contain several convolution terms. We propose a novel, efficient pseudo-spectral collocation scheme for computing the non-local terms in real space with the help of a specialized Gauss quadrature. Due to the exponential accuracy of the quadrature and a convenient choice of collocation points near interfaces, we can use grids with a significantly lower number of nodes than most other reported methods.
We demonstrate the capabilities of our numerical methodology by
studying equilibrium and dynamic two-dimensional test
cases with single- and multispecies hard-sphere and hard-disc particles
modelled with fundamental measure theory, with and without van der Waals
attractive forces, in bounded and unbounded physical domains. We show that our results satisfy statistical mechanical sum rules.
\end{abstract}




\section{Introduction}\label{Introduction}

Nanoscopic effects associated with fluid interfaces play an important role in a wide range of natural phenomena and technological applications. These include the design of water-repellent surfaces, inkjet printing, oil recovery, as well as a growing number of applications in the rapidly developing fields of micro- and nanofluidics \cite{Mitchell2001,Bonn.20090527,FrinkSalinger:2002:NumericalChallenges}. In the modern literature on soft-matter systems, particle-based approaches such as molecular dynamics (MD) or Monte-Carlo (MC) computations are popular tools which allow to obtain particle trajectories and phase-space configurations of statistical mechanical systems. However, the high numerical cost of direct simulations renders them intractable for large system sizes or long observation times.

In this context, classical density functional theory (DFT) has emerged in the past decades as a useful tool for investigations of nanoscale phenomena in soft-matter systems. DFT was first used to study fluid interfaces in the late 70's~\cite{Ebner:1977sw,Evans}, and since then has been applied to cover a wide range of fields~\cite{Wu:2007fh}, from the modeling of atomic~\cite{RothEvansLang:2002},
 molecular~\cite{Cao_Wu:2004} and polyatomic~\cite{ChandlerMcCoy:1986,YuWu:2002} systems to electrolyte solutions~\cite{Haertel:2015:DFTDoublelayer}, and even water interfaces~\cite{Hughes:2013:Water}. Within DFT, a fluid is described in terms of its one-body density alone.
This is achieved by approximating the free energy of the system as a functional of the density and obtaining the density profile from a minimisation procedure, e.g. by employing the variational principle \cite{Evans}. This is significantly less computationally demanding compared to MD-MC, whilst retaining nanoscale properties of the fluid \cite{Wu:2007fh}.

The typically employed DFTs for atomic fluids, such as a Lennard--Jones (LJ)
fluid, are based on the perturbation theory expansion of the free energy
density in powers of the attractive intermolecular potential, with the
reference free energy being that of a fluid with purely repulsive
intermolecular interactions, such as a fluid of hard spheres
\cite{Zwanzig:1954cq,TangWu:2003,Lasse:etal:2013}.  A highly accurate free
energy functional of a hard sphere fluid is provided by the
\revision{Fundamental Measure Theory (FMT) introduced by
Rosenfeld}~\cite{Rosenfeld:1989qc}. This employs a functional of
weighted densities, defined as convolutions of the fluid density with weight
functions of finite support. In the first order of the thermodynamic
perturbation theory, the attractive intermolecular interactions are typically
included via a mean-field approximation, as a convolution of the fluid
density with the algebraically decaying attractive part of the total
intermolecular potential.

The resulting integral equations are highly stiff and non-linear, and pose significant numerical challenges. Most commonly, the convolutions in these integral equations are solved by using fast Fourier transforms (FFT) \cite{Sears:2003uq,DissMerath:2008,Gross:2009:DFTSAFT,Knepley:2010,Zhou:2012three,Malijevsky:2013:CriticalPointWedgeFilling}.
This `matrix-free' approach has been employed in a variety of geometries in one- (1D), two- (2D) and three-dimensional (3D) domains (see Ref.~\cite{Sears:2003uq} for a comparison of computations in different dimensions).
A prerequisite for the use of FFT is a uniform cartesian grid~\cite{DissMerath:2008}, and accurate DFT computations therefore employ dense grids with 20~\cite{Malijevsky:2013:CriticalPointWedgeFilling} to 50~\cite{Gross:2009:DFTSAFT} discretisation points per hard core diameter to achieve adequate resolutions of the fluid density near interfaces. An improvement can be obtained if the weight functions, which exhibit discontinuities, are Fourier-transformed analytically~\cite{Knepley:2010}. This was also employed in Ref.~\cite{Zhou:2012three} in 3D with 10 discretisation points per hard sphere diameter. However, given the fact that far from the interfaces the fluid density is usually near-constant, such approaches are wasteful.
Another drawback is that methods based on FFT are restricted to the use of periodic boundary conditions. Non-periodic scenarios therefore have to be performed in large periodic domains mimicking non-periodicity, which means that jumps in the density profile are included in the domain.

Alternatively, a real-space finite element approach, discretised on a cartesian grid with linear interpolation, was introduced in Refs.~\cite{DouglasFrink:2000425,FrinkSalinger:2000kx,FrinkSalinger:2002:NumericalChallenges}. While this method is robust, it is also expensive~\cite{heroux:2007vn}.
Heroux~\etal~\cite{heroux:2007vn} reevaluated real-space approaches,
realising that the high computational cost for solving DFT problems was also
due to the use of preconditioners which are typically developed for partial
differential equations (PDEs) but not nonlocal ones such as DFT equations.
Therefore, an adapted solver was proposed in Ref.~\cite{heroux:2007vn} and applied to hard-sphere, polymer and molecular fluids. The respective code was made publicly available with the TRAMONTO software
package~\cite{TRAMONTO:2007}.

The aim of the present work is to introduce an alternative real-space quadrature based on the non-uniform pseudospectral discretisation, which can also be applied for unbounded physical domains. This allows the accurate discretisation of the fluid density profile with a small number of collocation points, by positioning collocation points close to fluid interfaces, therefore avoiding regions of near-constant fluid density. The convolutions are performed in real space, by separately discretising the intersection between the support of the density profile and the support of the respective weight functions, employing a quadrature scheme with spectral accuracy.

Our proposed scheme is highly accurate, efficient and fast. We have successfully applied it in a number of settings, for soft and hard-sphere fluids in 1D planar, spherically symmetric as well as 2D domains, for both equilibrium and non-equilibrium settings~\cite{Goddard:2012general,Goddard:2013Unification,Goddard:2013multi,Nold:FluidStructure:2014,nold2015nanoscale,Nold:2016:thesis}. A similar approach based on spectral methods was also used to compute the relaxation dynamics of planar films~\cite{YatsyshinSerafim:2012} and
equilibrium phase transitions in confinement~\cite{PeterPRE,PeterJChemPhys2,PeterJPhysCondMatt}.

The paper is organised as follows. Secs. \ref{sec:DFT} and \ref{sec:NumericalMethod} contain succinct reviews of DFT for LJ fluids and pseudospectral methods in unbounded domains, respectively. The numerical aspects associated with the discretisation and quadrature in 2D domains are detailed in Sec.~\ref{sec:SpectralTwoD}.
\revision{We verify our numerical method in Sec. \ref{sec:Validation} and validate} it with thermodynamic sum-rules in 1D and 2D settings, for single-fluid and multiple-species equilibrium and dynamic settings. \revision{Furthermore, we also include a} comparison with stochastic sampling techniques.
We conclude in Sec. \ref{sec:Conclusion} with the outlook and potential for further investigations.

\section{Model Equations \label{sec:DFT}}

Classical DFT is based on the theorem that the properties of an equilibrium
many-body system can be uniquely described by a functional of the number
density $\nDensity({\bf r})$~\cite{Evans,Wu-DFT,Mermin:1965fk}. This functional has two
main properties: (a) it is minimized by the equilibrium density distribution
and (b) at equilibrium it corresponds to the grand potential of the system.
For a single-component system, this functional is of the form
\begin{align}
\GrandPotential[\nDensity] = \FE[\nDensity] + \int \nDensity(\pos) \klammCurl{\Vext(\pos) - \chemPotN} \dI\pos,  \label{eq:GrandPotential}
\end{align}
where $\FE$ is the Helmholtz free energy functional, $\chemPotN$ is the chemical potential and $\Vext$ is the external potential of the system.
In order to obtain the equilibrium density distribution, the minimising condition (a) may be formulated by the application of the variational principle, leading to the Euler-Lagrange equation
\begin{align}
\frac{\delta \Omega[\nDensity]}{\delta \nDensity({\pos})} = 0.\label{eq:EulerLagrangeEquation}
\end{align}
The intrinsic Helmholtz free energy functional $\FE[\nDensity]$ can be split into an ideal-gas contribution, $\FEideal$, and an excess contribution due to the particle--particle interactions $\FEexc$
\begin{align}
\FE = \FEideal + \FEexc, \label{eq:FreeEnergyFunctional}
\end{align}
with
\begin{align}
\FEideal[\nDensity] = \beta^{-1} \int  \nDensity(\pos) \klamm{ \log \klamm{\Lambda^3 \nDensity(\pos) } - 1} \dI \pos,
\end{align}
where $\beta^{-1} = k_B T$, $k_B$ is the Boltzmann constant, $T$ the
temperature and $\Lambda$ the thermal wavelength. In contrast to the ideal-gas contribution, the excess free energy functional is not known exactly.
When modelling a Lennard--Jones fluid with particles interacting via
\begin{align}
\phiLJ(r) = 4 \LJdepth \klamm{ \klamm{\frac{\LJdiam}{r}}^{12} - \klamm{\frac{\LJdiam}{r}}^{6} }, \label{eq:LennardJonesPotential}
\end{align}
the excess free energy is typically split into contributions due to
short-range repulsive particle--particle interactions, $\FEhs$, and long-range attractive ones, $\FEattr$, which are treated in a perturbative
manner~\cite{Zwanzig:1954cq,TangWu:2003,Lasse:etal:2013}:
\begin{align}
\FEexc = \FEhs + \FEattr.
\end{align}
In \refe{eq:LennardJonesPotential}, $\LJdepth$ and $\LJdiam$ denote the
length and energy parameters of the Lennard--Jones potential, respectively.

As suggested by the notation, the repulsive contribution to the free energy
$\FEhs$ is usually approximated by a hard-sphere potential. One of the most
successful approaches to model hard-sphere fluids is \revision{FMT, first developed in
1989 by Rosenfeld}~\cite{Rosenfeld:1989qc,Roth:2010fk}, which is based on the geometrical properties of hard spheres~\cite{RosenfeldObituary:2002,RothEvansLang:2002}. Improvements to this original formulation have
been presented in order to capture the freezing
transition~\cite{rosenfeld1996dimensional,RosenfeldTarazona:1997,Tarazona:2000:Freezing}, or a more accurate equation of state~\cite{Lang:2001:thesis,RothEvansLang:2002,YuWu:2002:Modified}. These
modified formulations, however, lead to equations with a functional form
close to those of the original formulation, which is still widely used
\cite{Malijevsky:2013:CriticalPointWedgeFilling,Gor:2012:QuenchedSolid,ArcherEvans:2013,Knepley:2010,RosenfeldObituary:2002}
and gives very accurate results if compared with MD or MC computations
\cite{Roth:2010fk}. In the following, for the sake of simplicity we restrict ourselves to the implementation of the original \revision{FMT by Rosenfeld}, and briefly discuss the formalism for 3D hard spheres and 2D hard disks. But, it should be noted, that our numerical framework can be easily adapted to other FMT formalisms.

\subsection{Hard-sphere FMT}
For a system of multiple species of hard spheres with radii $R_i$ and density $\nDensity^{(i)}$,
Rosenfeld formulated the hard-sphere free energy as a functional of scalar- and vector-weighted densities $\nDensity_\alpha$, given by
\begin{align}
\FEhs[\{\nDensity^{(i)}\}] &= \beta^{-1} \int \UpsHS \klamm{\klammCurl{\nDensity_\alpha\klamm{\pos}}} \dI \pos, \label{eq:DefRFMT_HardSphereContribution1}
\end{align}
with
\begin{align}
\UpsHS \klamm{\{\nDensity_\alpha\}} &\defi
- \nDensity_0 \ln \klamm{1-\nDensity_3} + \frac{\nDensity_1 \nDensity_2}{1-\nDensity_3} + \frac{\nDensity_2^3}{24\pi \klamm{1-\nDensity_3}^2}
- \frac{{\bf \nDensity}_1\cdot{\bf \nDensity}_2}{1-\nDensity_3} - \frac{\nDensity_2\klamm{{\bf \nDensity}_2 \cdot {\bf \nDensity}_2}}{8\pi \klamm{1-\nDensity_3}^2}
. \label{eq:DefRFMT_HardSphereContribution2}
\end{align}
The weighted densities $\nDensity_\alpha$ are defined through convolutions of weight functions $\weight_{\alpha,i}$ with the density field $\nDensity$:
\begin{align}
\nDensity_\alpha \klamm{\bf r} \defi
\sum_i \klamm{\nDensity^{(i)} \ast \weight_{\alpha,i}}\klamm{\pos}
= \sum_i \int \nDensity^{(i)}\klamm{\pos '} \weight_{\alpha,i} \klamm{\pos - \pos'} \dI \pos'.\label{eq:nAlphaDef}
\end{align}
The vector-weighted densities ${\bf \nDensity}_\alpha$ are defined analogous to Eq.~\refe{eq:nAlphaDef}, with vector weights ${\boldsymbol \weight}_{\alpha,i}$. The weights, as obtained through the decomposition of the Mayer function~\cite{Rosenfeld:1989qc}, are
\begin{align}
\weight_{2,i}\ofR \defi \delta\klamm{R_i - |\pos|},\qquad
\weight_{3,i}\ofR \defi \Theta\klamm{R_i - |\pos|},\qquad
{\boldsymbol \weight}_{2,i}\ofR \defi \frac{\pos}{|\pos|} \delta\klamm{ R_i - |\pos| }, \label{eq:WeightedDensitiesRFMT2}
\end{align}
where $\delta(x)$ is the Dirac delta and $\Theta(x)$ is the Heaviside step function.
The remaining weights $\weight_{1},\weight_0,{\boldsymbol \weight}_1$ depend linearly on the weights defined
in (\ref{eq:WeightedDensitiesRFMT2}) and are defined by
\begin{align}
\weight_{1,i}\ofR \defi \frac{\weight_{2,i}\ofR}{4\pi R_i} ,\qquad
\weight_{0,i}\ofR \defi \frac{\weight_{2,i}\ofR}{4 \pi R_i^2} ,\qquad
{\boldsymbol \weight}_{1,i}\ofR \defi \frac{{\boldsymbol \weight}_2\ofR}{4 \pi R_i}.
\end{align}
We note that the system may be reduced by making use of the fact that the vector-weighted density ${\bf n}_2$ is the negative gradient of $n_3$: ${\bf n}_2 = -\Grad \nDensity_3$, such as applied
by Merath~\cite{DissMerath:2008}.
Insertion of Eq.~(\ref{eq:DefRFMT_HardSphereContribution1}) into the
Euler-Lagrange equation \refe{eq:EulerLagrangeEquation}, leads to a
contribution due to hard-sphere effects of
\begin{align}
\frac{\delta \FEhs[n]}{\delta \nDensity^{(i)}(\pos)} = \beta^{-1} \sum_{\alpha} \klamm{\diff{\UpsHS}{\nDensity_{\alpha}} \ast \hat \weight_{\alpha,i}}\klamm{\pos},
\label{eq:Rosenfeld_Variational}
\end{align}
where $\hat \weight_{\alpha,i}\klamm{\pos} = \weight_{\alpha,i}\klamm{-\pos}$.
The corresponding 2D convolution weight for a system which is invariant in one direction is given in Sec.~\ref{sec:ConvWeights2D} of the Appendix.

\subsection{Hard-disk FMT \label{sec:DiskFMT}}
An analogous approach can be followed if hard disks are considered. Here, we follow the description by Roth \etal~\cite{RothMeckeOettel2012} and define the hard-disk free energy density as
\begin{align}
\UpsHD \klamm{\{\nDensity_\alpha\}} \defi &- \nDensity_0^\HD \ln \klamm{1-\nDensity_3^{\HD}}+\\
& \frac{1}{4\pi \klamm{1-\nDensity_3^{\HD}}}\klamm{ \frac{19}{12} \klamm{\nDensity_2^{\HD}}^2  - \frac{5}{12} {\bf \nDensity}_2^{(1,\HD)}\cdot{\bf \nDensity}_2^{(1,\HD)} - \frac{7}{6} {\bf \nDensity}_2^{(2,\HD)} \cdot {\bf \nDensity}_2^{(2,\HD)} }. \notag
\end{align}
The weighted densities are defined analogously to (\ref{eq:nAlphaDef}) in 2D:
\begin{align}
\weight_{2,i}\ofR \defi \delta\klamm{R_i - |\pos|}
\qquad\text{and}\qquad
\weight_{3,i}\ofR \defi \Theta\klamm{R_i - |\pos|}. \label{eq:w2w3:HD}
\end{align}
Additional weights are
\begin{align}
\weight_{0,i}^\HD\klamm{\pos} = \frac{\weight_{2,i}\klamm{\pos}}{2\pi R_i},\qquad
{\boldsymbol \weight}_{2,i}^{(1,\HD)}\klamm{\pos} = {\bf \pos} \delta\klamm{R_i - |\pos|},\qquad
{\boldsymbol \weight}_{2,i}^{(2,\HD)}\klamm{\pos} = \klamm{{\pos} \otimes {\pos}} \delta\klamm{R_i - |\pos|}.
\end{align}
We note that for the hard-disk case, the convolutions defining the weighted densities are defined in $\mathbb{R}^2$, as opposed to the weighted densities employed for the hard-sphere free energy, which are defined in $\mathbb{R}^3$.

\subsection{Attractive free energy contribution}

We follow the route of treating the attractive forces as perturbations to the repulsive forces~\cite{Lasse:etal:2013},
by including them in a mean-field manner as
\begin{align}
\FEattr[\nDensity] &= \frac{1}{2 } \iint \BHattr({|\pos - \pos'|}) \nDensity(\pos)\nDensity(\pos') \dI\pos' \dI\pos, \label{eq:FEattr}
\end{align}
where $\BHattr$ represents the attractive particle--particle interactions, modeled here with
\begin{align}
\BHattr\klamm{r} =
\depthLJ \left\{ \begin{array}{ll}
0 & \text{for } r \leq  \LJdiam\\
4 \klamm{
\klamm{\displaystyle\frac{\LJdiam}{r}}^{12} -
\klamm{\displaystyle\frac{\LJdiam}{r}}^{6}
} & \text{for } r > \LJdiam
\end{array}  \right. .\label{eq:pattr}
\end{align}
The contribution of the free energy functional $\FEattr$ to the Euler-Lagrange equation (\ref{eq:EulerLagrangeEquation}) is
\begin{align}
\frac{\delta \FEattr[\nDensity]}{\delta \nDensity({\pos})}
= \klamm{\phi_{\text{attr}} \ast \nDensity }\klamm{\pos} = \int \BHattr({|\pos - \pos'|}) \nDensity(\pos') \dI\pos'. \label{eq:AttrContrEulerLagrange}
\end{align}
The corresponding 2D convolution weight for a system which is invariant in one direction is given in Sec.~\ref{sec:ConvWeights2D} of the Appendix.

\subsection{Dynamic DFT \label{DDFT}}

An analogous result to that underpinning DFT holds for the dynamics of many-body systems \cite{Chan:2005uq},
leading to a dynamic DFT or DDFT.  DDFTs typically take the form of a continuity equation
\[
	\partial_t \nDensity(\pos,t) = \grad{\pos} \cdot \big( \mathbf{J}([\nDensity(\pos,t)],\pos,t) \big),
\]
where $t$ denotes the time and the challenge is to determine the flux $\mathbf{J}$, which is a functional of the density.  Due to the success
of DFT in describing nanoscale fluid properties, this functional is usually based on the free energy of a related equilibrium system.  In particular, this ensures
that the DDFT reduces to the correct DFT at equilibrium.  In this way, DDFT can be regarded as a natural generalisation
of DFT.

The first DDFTs for colloidal particles were obtained phenomenologically~\cite{Evans,DietrichFrisch:1990}, but there have since been
a number of attempts to rigorously derive an overdamped/high friction DDFT from the Smoluchowski equation for the
full $N$-body density.  Marconi and Tarazona \cite{Marconi:1999ys,Marconi:2000ys} derived the first DDFT for pairwise interparticle potentials,
\begin{align}
	\partial_t \nDensity(\pos) =
	\frac{1}{\frictionCoeff m}
	\grad{\pos} \cdot \Big[ \nDensity(\pos,t) \grad{\pos}
	\frac{\delta \FE [\nDensity(\cdot,t)]}{\delta \nDensity(\pos,t)} \Big],
\end{align}
where $m$ is the mass of one particle, $\frictionCoeff$ is the friction
coefficient of the system and $\FE$ is the free energy functional for a
corresponding equilibrium system with the same one-body density, such as that
given in (\ref{eq:FreeEnergyFunctional}), but including the external
potential $\Vext$. This additional approximation is often called the
adiabatic approximation.  Later, this result was generalised to $N$-body
interactions~\cite{Archer:2004mb}. Hydrodynamic interactions, caused by the
interplay between the flow in the bath and the colloidal movement, have also
been included in more advanced DDFT
models~\cite{Rex:2009qf,BenDDFT,Goddard:2012general,Goddard:2013Unification}.

In order to incorporate the effect of (microscopic) inertia in DDFT, two
different approaches have been considered. The first uses a multiple
timescale technique
\cite{Marconi:2007zr,Marconi:2006zr,tarazona2008beyond,MaroniTarazona:2008:Beyond},
whilst the second takes momentum moments of the Kramers equation (the
underdamped analogue of the Smoluchowski equation), leading to a continuity
equation coupled to an evolution equation for the
velocity~\cite{Archer.20090107},
\begin{align}
	\partial_t \nDensity(\pos,t) &= - \grad{\pos} \cdot \big( \nDensity(\pos,t) \mathbf{v}(\pos,t) \big) \\
	\partial_t  \mathbf{v}(\pos,t) &= -  \mathbf{v}(\pos,t) \cdot \grad{\pos} \mathbf{v}(\pos,t)
	- \gamma  \mathbf{v}(\pos,t)
	- \frac{1}{m} \grad{\pos} \frac{\delta \FE [\nDensity(\pos,t)]}{\delta \nDensity(\pos,t)}, \label{eq:DDFT:MomentumEq}
\end{align}
where $\mathbf{v}$ is the velocity distribution.  This formalism has also recently
been extended to include hydrodynamic interactions~\cite{BenDDFT,Goddard:2012general,Goddard:2013Unification}.
To close the infinite hierarchy of moment equations, one typically utilises
the local equilibrium approximation in which the momentum distribution is
assumed to be locally Maxwellian. DDFT has been successfully applied to a
wide variety of problems including sedimentation, cluster formation and cell
modelling, with results having been validated against the full underlying
stochastic dynamics
\cite{Rex:2009qf,Goddard:2012general,Goddard:2013Unification}.

\subsection{Nondimensionalization}

We nondimensionalise the system with a length and an energy scale.
For the sake of simplicity, we only consider hard-sphere or hard-disk mixtures where particles of all species have the same radius $R_i = R$. In particular, we take the length-scale
of the Lennard--Jones potential $\LJdiam$ to be equal to the hard-sphere diameter $2R$, defining the length scale. As an energy scale for systems with interparticle attraction, we employ $\eD$ such that
\begin{align}
-\frac{32}{9} \pi \eD = \int_{\mathbb{R}^3} \BHattr\klamm{|\pos|} \dI \pos. \label{eq:defineEnergyScale}
\end{align}
For an attractive particle--particle interaction \refe{eq:pattr} without radial cutoff, we get $\depthLJ_D = \depthLJ$. If, however, a cutoff radius $r_c$ for the attractive interaction potential is employed for computational purposes, then
\refe{eq:defineEnergyScale} ensures that the fluids are represented by the same dimensionless bulk phase diagram, independent of the cutoff radius $r_c$, hence improving comparability.
For pure hard-sphere systems, we set $\eD = k_BT$.
We thus introduce the following dimensionless quantities
\begin{align}
\tilde \pos &= \frac{\pos}{\sigma},\quad
\tilde T = \frac{k_B T}{\eD},\quad
\tilde \phi_{\text{attr}} = \frac{\BHattr}{\eD},\quad
\tilde \nDensity = \nDensity \LJdiam^3,\notag\\
\tilde \chemPotN &= \frac{\chemPotN}{\eD} - \frac{k_B T}{\eD} \log \klamm{\frac{\Lambda^3}{\LJdiam^3}} \quad \text{and} \quad
\tilde V_{\text{ext}} = \frac{\Vext}{\eD}.
\end{align}
The second term in the nondimensionalisation of the chemical potential accounts for the contribution of the term $\Lambda$ in the ideal-gas free energy and
leads to a constant shift of the dimensionless chemical potential.

For the dynamic computations, we introduce different time-scales for the overdamped and the inertial case, given by
\begin{align}
T_{\text{OD}} = \frac{\frictionCoeff m \LJdiam^2}{\eD}
\qquad \text{and}\qquad
T_{\text{IN}} =\LJdiam \sqrt{\frac{m}{\eD}},
\end{align}
respectively. The dimensionless friction coefficient in the momentum equation (\ref{eq:DDFT:MomentumEq}) then becomes
\begin{align}
\DL \frictionCoeff = \frictionCoeff \LJdiam \sqrt{\frac{m}{\eD}}.
\end{align}
In what follows, we omit the tildes to simplify our notation.

\newcommand{\FT}{\mathcal{F}}

\section{Numerical Method \label{sec:NumericalMethod}}

The accurate computation of convolutions
\begin{align}
(\nDensity \ast \convWeight)(y) \defi \int \convWeight(y- \tilde y) \nDensity(\tilde y) \dI \tilde y
	\label{convolutionDefn}
\end{align}
 is crucial for solving the
Euler-Lagrange equation \refe{eq:EulerLagrangeEquation} and ultimately
obtaining a reliable equilibrium density profile. Convolutions are needed to
compute the weighted densities as given in Eq.~\refe{eq:nAlphaDef}, for the
hard-sphere contribution to the free energy (see
Eq.~\refe{eq:Rosenfeld_Variational}) as well as the attractive contribution
to the free energy (see Eq.~\refe{eq:AttrContrEulerLagrange}). The obvious
choice is to employ the convolution theorem, and perform the computation of
the convolutions in Fourier space~\cite{Roth:2010fk} through
\begin{align}
\nDensity \ast \convWeight = \FT^{-1}\klammCurl{ \FT\klammCurl{\nDensity} \cdot \FT\klammCurl{\convWeight}  }, \label{eq:ConvolutionFourier}
\end{align}
where $\FT$ is the Fourier transform. Employing FFT is of $O(N \log N)$ complexity, where $N$ is the number of Fourier
modes. 

Note, however, that a density profile of a liquid film adsorbed on a substrate exhibits a jump at the wall, and approaches continuously its bulk value with increasing distance to the wall. Meanwhile, any decomposition in Fourier modes requires a truncation of the
physical domain and the assumption of periodicity. This entails two crucial difficulties: First, it requires a special treatment of the Gibbs phenomenon induced by the discontinuity of the density profile at the wall. Similarly, the representation of weight functions $\weight$ for FMT such as given in Eqs.\refe{eq:WeightedDensitiesRFMT2} and \refe{eq:w2w3:HD} needs special treatment due to the short-range finite support properties of these functions~\cite{Roth:2010fk}. Second, non-periodic scenarios have to be mimicked employing large periodic domains, therefore wasting precious computation power.

Several approaches have been presented to circumvent these difficulties. By
computing the Fourier transform analytically, the accuracy may be improved
substantially~\cite{Knepley:2010}. Alternatively, expansions of the
equilibrium density profile in terms of the local curvatures at the wall have
been proposed~\cite{konig:2005curvature}. Here, however, we aim to avoid the complications introduced by computing the convolutions in Fourier space
altogether by performing the convolutions in real-space. This approach, which
we describe below, is based on pseudospectral methods.

\subsection{Chebyshev pseudospectral method}

Here we outline a real-space discretisation method for FMT-DFT employing a
pseudospectral method~\cite{Boyd:2001Cheb}. In this framework, a function
$f(x)$ defined on $x\in[-1,1]$ is represented by its functional
values at the collocation points $x_n$. The function is then approximated by the interpolating polynomial, defined via
\begin{align} 
p_N(x)\defi \sum_{n=0}^N f(x_n) P_n(x),
\label{eq:Decomposition} 
\end{align} 
where $P_n(x)$ denote the Lagrange polynomials of degree N,
given by
\begin{align}
P_n(x) \defi  \prod_{m=0,m\neq n}^{N}{ \frac{ x-x_m}{x_n-x_m}},
\label{eq:Legendre}
\end{align}
with the property
\begin{align}
\revision{
P_n\klamm{x_k} = \left\{ 
\begin{array}{ll}
1 & \text{for } k = n\\
0 & \text{for } k \neq n
\end{array}
\right. .}
\end{align} 
To avoid the so-called Runge phenomenon that occurs in polynomial
interpolation over equispaced grids, we choose the so-called Gauss-Lobatto-Chebyshev collocation points
\begin{align}
x_{n} &\defi \cos t_n 
\qquad \text{with}\qquad
t_n \defi \frac{\pi n}{N} 
 \qquad \text{for}\qquad n = 0\ldots N, \label{eq:Spectral:ChebCollPts}
\end{align}
which are clustered at the endpoints of the interval. By doing
so, we obtain a spectrally accurate representation of any smooth function $f(x)$ over the whole interval. 
Using Eq.~\refe{eq:Decomposition}, the value of $p_N(x)$ can, in principle, be
computed anywhere in the domain. Nevertheless, it is generally known that the calculation of Eq.~\refe{eq:Legendre} is both costly and numerically unstable. These issues can be avoided by using the barycentric formula~\cite{Berrut:2004ly,Tee:2006Ad}
\begin{align}
p_N(x) = {\displaystyle\frac{{\mathlarger\sum_{k=0}^N} {\displaystyle\frac{\bar w_k}{x - x_k}f(x_k) }}{{\mathlarger\sum_{k=0}^N}  {\displaystyle\frac{\bar w_k}{x-x_k}}}}, \label{eq:NumericalMethods:BarycentricForm}
\end{align}
where ${\bar w}_k$ are the so-called barycentric weights, which, for the Gauss-Lobatto-Chebyshev points are given by~\cite{Salzer:1972}
\begin{align}
\bar w_j \defi \klamm{-1}^j d_j\qquad \text{with}\qquad
d_j \defi \left\{ 
\begin{array}{ll}
1/2 & \text{for } j\in \{0,N\} \\
1 &\text{otherwise}
\end{array}
\right. .\label{eq:WeightBaryIP}
\end{align}
The integral of $f$ over $[-1,1]$ is computed through the use of the Clenshaw-Curtis quadrature~\cite{clenshaw:1960}
\begin{align}
\int_{-1}^1 f(x) dx = \sum_{k=0}^N w_k f(x_k),\label{eq:Spectral:Clenshaw:Quadrature}
\end{align}
with weights
\begin{align}
w_j \defi \frac{2d_j}{N} \left\{
\begin{array}{ll}
 1 - {\mathlarger\sum_{k=1}^{(N-2)/2}} \fracd{ 2\cos\klamm{2kt_j}}{4k^2-1} - \fracd{\cos\klamm{\pi j}}{N^2-1}  &\qquad \text{for } N \text{ even}\\
 1 - {\mathlarger\sum_{k=1}^{(N-1)/2}} \fracd{ 2\cos\klamm{2kt_j}}{4k^2-1}&\qquad \text{for } N \text{ odd}
\end{array}\right..
\label{eq:Num:Cheb:ClenshawCurtisQuad1}
\end{align}

\subsection{Unbounded domains \label{sec:Spectral:UnboundedDomains}}
When dealing with functions approaching a constant value in an unbounded
domain (such as the density profile of a fluid in contact with a wall -- the
constant value being approached far from the wall) several numerical
techniques can be used, such as~\cite{Boyd:2001Cheb,Shen2009}:
\begin{enumerate}[(1)]
\item domain truncation,
\item approximation by orthogonal systems in the unbounded domain, or
\item mapping of the unbounded domain onto a bounded domain and applying standard spectral methods.
\end{enumerate}
Option (1) is appropriate when dealing with quickly-decaying functions. It has
been widely used in DFT computations. As far as options (2) and (3)
are concerned, the main difference is that in (2) the equations are studied
in the physical (unbounded) domain, while in (3) the equations are considered
in the mapped bounded domain, which is subsequently referred to as the
computational domain. Here, we adopt (3) as its implementation provides a
greater flexibility in studying different domain geometries. In other words, once
the basic methods are implemented in the computational domain, we only have
to adapt the mappings to the physical domain to study bounded, unbounded
domains and combinations of both in 2D. It should also be emphasised that we have already successfully applied (3) in several physical settings, see Refs.~\cite{Goddard:2013multi,Goddard:2012general,Goddard:2013Unification,Nold:FluidStructure:2014,nold2015nanoscale,YatsyshinSerafim:2012,PeterPRE,PeterJPhysCondMatt,PeterJChemPhys2}.

\begin{figure}[h]
\centering
	\includegraphics[width=8cm]{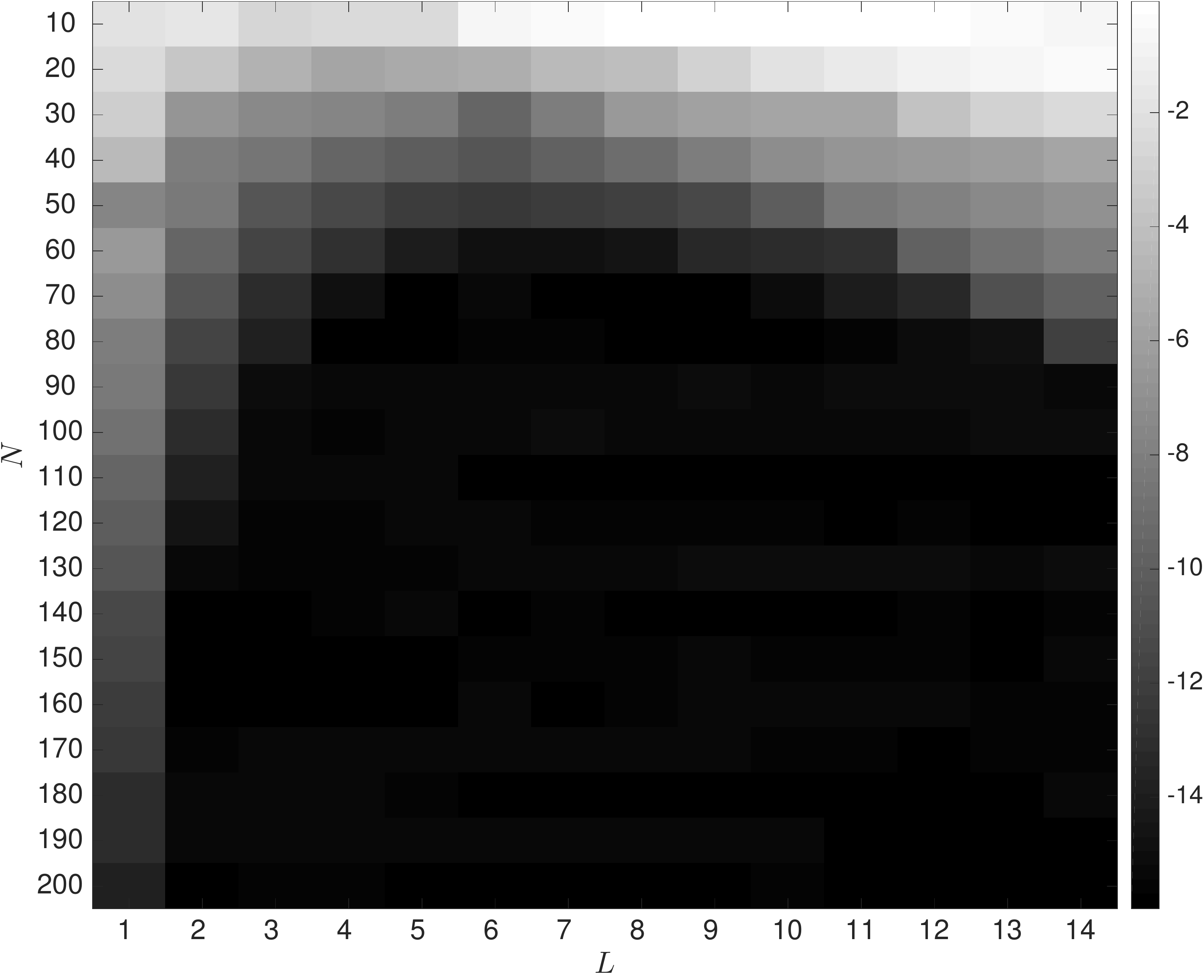}
	\caption{\label{fig:LSensitity} Behavior of the $\log_{10}$-logarithm of the relative error for the integral of the test function
    $f(y) = (2/\pi)^{1/2} y^2 \exp\klamm{-\frac{y^2}{2}}$, given a discretisation
    of the whole line with mapped Gauss-Lobatto-Chebyshev collocation points using the algebraic map
    $\AlgMapWhole$ in Eq.~\refe{eq:WholeLineAlgebraicMapping}, as a function of
    the number of collocation points and the mapping parameter $L_1$. The error is given in a logarithmic grayscale from white (corresponding to $10^0$) to black (corresponding to  $10^{-15})$.}
\end{figure}

We distinguish between algebraic, logarithmic and exponential maps from the
computational domain $[-1,1]$ to the semi-infinite and infinite domains
$[0,\infty]$ and $[-\infty,\infty]$, respectively. These have been compared
extensively in~\cite{Boyd:2001Cheb}, as well as in the earlier
studies~\cite{Grosch:1977273,Boyd:198243}. In a minimalist approach, we
choose the algebraic maps taking advantage of their robustness. In the
direction normal and parallel to the wall, i.e.\ for a semi-infinite and
infinite physical domains, the algebraic maps are
\begin{align}
\AlgMapWhole &: [-1,1]\to [-\infty,\infty] \quad x \to L_1 \frac{x}{\sqrt{1-x^2}}, \label{eq:WholeLineAlgebraicMapping}
\end{align}
and
\begin{align}
\AlgMapSemi &: [-1,1]\to [0,\infty] \quad x \to L_2 \frac{1+x}{1-x}, \label{eq:SemiInfiniteAlgMapping}
\end{align}
respectively. $L_{1,2}$ are mapping parameters linked to the length scale
of the physical domain. In particular, $50\%$ of the discretisation points of
the computational domain $[-1,1]$ are mapped onto the intervals $[-L_1,L_1]$
and $[0,L_2]$ in the physical domain, respectively.
One further advantage of using algebraic maps is that they retain their optimal properties as $N\to
\infty$ for constant mapping parameters $L$~\cite{Boyd:2001Cheb}. In
Fig.~\ref{fig:LSensitity}, the low sensitivity of the algebraic map
$\AlgMapWhole$ with respect to $L_1$ is demonstrated by varying the number of collocation points, and comparing an analytical expression for the integral of a test function with the numerical result (see also Ref.~\cite{Boyd:198243}). Based on this, the parameters $L_1,L_2$ are chosen empirically by taking into consideration the expected typical length scales of the density profile.

\subsection{Computation of convolutions \label{S:convolution}}

We compute the convolution of the density profile $\nDensity$ and a weight function $\convWeight$ in real space. If the density profile is represented on a bounded domain by a relatively fine mesh, then the weight function $\convWeight$ may be represented on the same grid as the density distribution $\nDensity$
and the computation of Eq.~\refe{convolutionDefn} is straightforward.

\begin{figure}[htbp]
\centering
\includegraphics{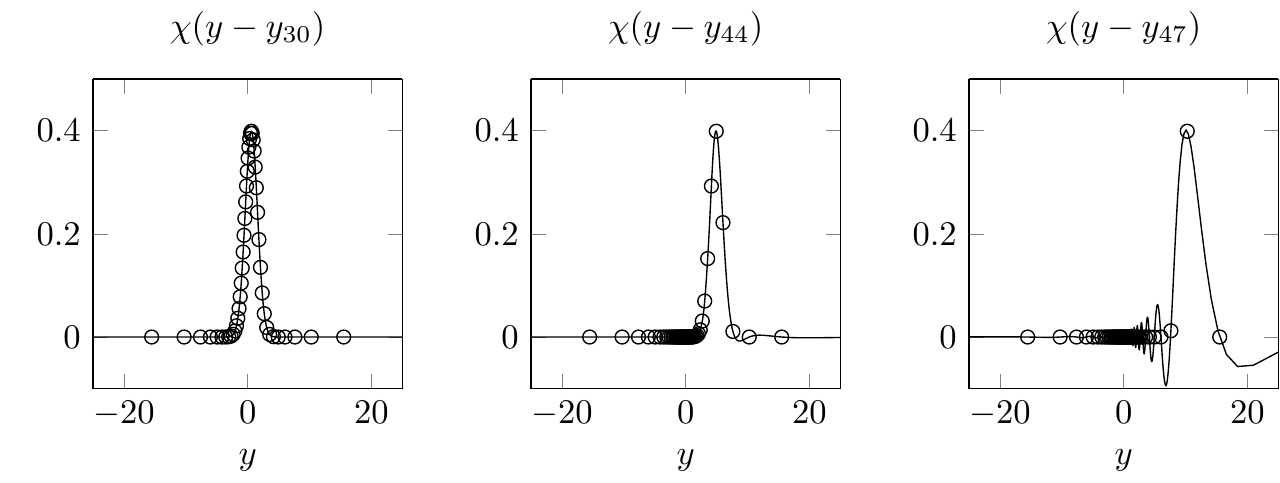} 
\caption{Interpolation of three Gaussian distributions $\convWeight(y) = \frac{1}{2\pi}e^{-y^2/2}$, shifted by the position of
the $30^{\text{th}}$, $44^{\text{th}}$, and $47^{\text{th}}$ collocation
points for the left, middle and right subplot, respectively. The whole line is
discretized with $50$ Gauss-Lobatto-Chebyshev collocation points mapped from $[-1,1]$ to
$[-\infty,\infty]$ using the algebraic map $\AlgMapWhole$ in Eq.~\refe{eq:WholeLineAlgebraicMapping} with mapping parameter $L_1 = 2$.}
\label{Fig:kernelAll}
\end{figure}

Typically, however, we wish to compute the convolution of a relatively narrow
function $\convWeight$, with support of the order of the particle diameter, with a
density distribution $\nDensity$ which varies smoothly over a large domain.
These density distributions can be accurately represented with a relatively
small number of collocation points on the full domain. However, if one tries
to represent the weight function $\convWeight$, shifted by the position of the
collocation point $y_k$, on the same set of collocation points as the density
profile, the result can be highly inaccurate. To illustrate this behavior, we plot
in Fig.~\ref{Fig:kernelAll} the interpolation of three shifted Gaussian
distributions on the same grid centred at $y=0$. It can clearly be seen how
the quality of the representation rapidly deteriorates as the Gaussian is
shifted to positions at which the grid becomes coarse.

In order to account for this behavior, we rewrite the convolution \refe{convolutionDefn} as
\begin{align}
(\nDensity \ast \convWeight)(y) = \int_{\Intersection(y)} \convWeight(- \tilde y) \nDensity_y(\tilde y) \dI \tilde y, \label{eq:ModifiedConvolutionExpression}
\end{align}
with the shifted density distribution 
\begin{align}
\nDensity_y(\tilde y) \defi \nDensity( y + \tilde y) \label{eq:Numerics:DefineShiftedDensity}
\end{align}
and where $\Intersection(y)$ is the intersection between the support of the weight function $\convWeight$ and the support of the shifted density distribution $\nDensity_y$, similar to the procedure employed in Ref.~\cite{elGendi:1969} in 1D. Assuming that the density profile changes slowly, we conclude that the behavior of the integrand in (\ref{eq:ModifiedConvolutionExpression}) is dominated by the weight $\convWeight$. We therefore discretise the domain $\Intersection(y)$ for each collocation point $y = y_k$ based on the behavior of $\convWeight$, and interpolate the shifted density profile $\nDensity_y$ onto this grid.

In particular, we construct a convolution matrix $\ConvOp$ which acts on the vector $\nDensity_k$ of values of $\nDensity$ at the collocation points $y_k$ and returns $\klamm{\convWeight \ast \nDensity}_k$ at the collocation points $y_k$. The $k$-th row of $\ConvOp$, $\klamm{\ConvOp}_k$, is then computed as follows:
\begin{enumerate}
\item Discretise $\Intersection(y_k)$ with $M$ collocation points $\hat y_M$ and compute the integration weights $\hat w_M$ for this domain. [row vector, length $M$].
\item Evaluate the weight function $\convWeight_M = \convWeight(-\hat y_M)$ on the
    discretisation points of $\Intersection(y_k)$ [column vector, length
    $M$].
\item Using \refe{eq:NumericalMethods:BarycentricForm}, compute the interpolation matrix ${\bf IP}_k$ of the original grid $y_N$ onto the grid of $\Intersection(y_k)$, given by $\hat y_M$, and computed in step 1.
[$M \times N$ matrix].
\item Set $\klamm{\ConvOp}_k = \revision{\hat w_M} \text{diag}\klamm{\convWeight_M} {\bf IP}_k$.
\end{enumerate}
Here, $\text{diag}(v)$ is the square matrix with the elements of the vector $v$ on the diagonal and zeros elsewhere.

This procedure apparently has a much higher complexity --- determined by the
computation of the interpolation matrices --- of $O\klamm{MN^2}$, if compared
to the computation of the convolution in Fourier space, which is of $O(N\log
N)$ complexity. \revision{We note that this computation of the interpolation matrices is
done point-by-point and is therefore easily parallelisable.}

\revision{Also,} due to its flexibility in describing the domain
$\Intersection(y_k)$, the procedure described here naturally enables the accurate
description of weight functions and density profiles with discontinuities and
finite support. This allows us to use a smaller number of discretisation
points for both the discretisation of the density profile and of
$\Intersection(y_k)$.

\revision{Another way to bring down the complexity of the computation of the interpolation matrices, is to only
consider points that are close to $y_k$. This use of a cutoff for the computation of the interpolation matrices does speed up computations, but the speed-up is not expected to be as dramatic because of the clustering of the points near high-gradient regions. Also, the introduction of such a cutoff will affect convergence properties and therefore has to be studied carefully. For this reason, we opted not to utilise such a cutoff in this work.}

\revision{It is noteworthy that the procedure only has
to be performed once for each geometry. In particular,} in the process of solving the Euler-Lagrange equation
\refe{eq:EulerLagrangeEquation}, computing the convolution reduces to a
matrix-vector multiplication of complexity $O(N^2)$. We emphasise that a direct comparison of numerical procedures by means of their complexity is meaningful only if the number of points for both procedures is sufficiently large, of the same
order of magnitude, and the results are of the same accuracy. We show in the
following that with the real-space convolution procedure described here, a higher
accuracy can be achieved while employing a considerably lower number of
discretisation points, hence trading off its higher complexity. 

\subsection{Pseudospectral methods in 2D domains \label{sec:SpectralTwoD}}

The procedure is now generalised to 2D by
appropriate application of tensor products~\cite{Trefethen_2000}, such as done e.g. in
\cite{Le:1992} to solve the Navier--Stokes (NS) equations or in
\cite{Mai:2007} to solve 2D biharmonic boundary value problems. In
particular, the vector space of 2D polynomials approximating a function 
$f\klamm{\bf x}$ defined on the unit cell $[-1,1]\times[-1,1]$, which here 
is the computational domain in 2D, is defined as the tensor product of the 1D
polynomial vector spaces ${\bf P}_{N_1} \otimes {\bf P}_{N_2}$. Here, $N_1, N_2$ are the
number of collocation points along the first and second axes, respectively.
The unit cell $[-1,1] \times [-1,1]$ is discretised with sets of 1D collocation points
${\bf x}_{1,2}\in [-1,1]^{N_{1,2}}$ in each direction, defined in Eq.~\refe{eq:Spectral:ChebCollPts}~\cite{Trefethen_2000}. Details are given in Appendix \ref{sec:NumericalRepresentation}.

In Sec.~\ref{sec:Spectral:UnboundedDomains}, we have mapped the 1D
computational domain $[-1,1]$ onto the whole line $[-\infty,\infty]$ and a
semi-infinite space $[0,\infty]$, employing the algebraic maps $\AlgMapWhole$
and $\AlgMapSemi$, respectively. Here, we proceed analogously for 2D and map
the computational domain onto the physical domain through a general bijective
map
\begin{align}
\PhysSpace: (x_1,x_2) \to (y_1',y_2') = \klamm{\PhysSpace_1(x_1,x_2),\PhysSpace_2(x_1,x_2)}. \label{eq:General2DMapFromCompDomain}
\end{align}
Here, $\PhysSpace_{1,2}$ are scalar functions which correspond, depending on the physical domain we wish to discretize, to $\AlgMapWhole,\AlgMapSemi$ or to a simple scaling function.
Depending on the geometry, the first and the second variables of the
computational domain may be discretised employing either Gauss-Lobatto-Chebyshev collocation points in the interval
$[-1,1]$ or an equispaced grid on $[0,1)$ for periodic functions. 

The variables $(y_1',y_2')$ in the physical domain are then mapped to the Cartesian domain
via 
\begin{align}
\CartMap: (y_1',y_2') \to (y_{1},y_{2}). \label{eq:DefCartMap}
\end{align}
In particular, depending on the geometry we wish to discretise, the physical domain may be 
represented in polar, spherical or skewed cartesian coordinates. The maps $\CartMap$ for these
cases are given in Sec.~\ref{sec:App:RepPhys} of the Appendix.

\subsection{Domain discretisation for DFT-FMT equations in the half-space \label{sec:DomainDiscr}}

Modeling a fluid with FMT requires the discretisation not only of the density
$\nDensity$, but also of the weighted densities defined through
Eq.~(\ref{eq:nAlphaDef}). If the domain of the particles is confined by a
hard wall, we obtain $\nDensity |_{y_2<0} = 0$. Considering a hard-sphere
fluid with spheres of diameter one, and employing the weights defined in
Eq.~(\ref{eq:WeightedDensitiesRFMT2}), then it becomes clear that the support
of the weighted densities is $\{(y_1,y_2):y_2 > -1/2\}$. Hence, a separate
discretisation of the weighted densities is necessary.

Furthermore, if the contact density of a hard-sphere fluid at the wall is greater than zero,
then the derivatives of the weighted densities in the direction normal to the
wall, $\diff{\nDensity_{\alpha}}{y}$ exhibit a jump at $y_2= 1/2$. This can
best be seen for the behavior of $\nDensity_2$ and the second component of
$\bf \nDensity_2$ in the computation of a hard-sphere fluid in contact with
the hard wall, such as depicted in Fig.\ \ref{fig:Density_DFTFMT}. There, the
weighted densities $\nDensity_2$ and $\bf \nDensity_2$ are clearly
non-differentiable at $y_{2} = 1/2$.

\begin{figure}
	\centering
		\begin{subfigure}[t]{0.47\textwidth}
		\includegraphics{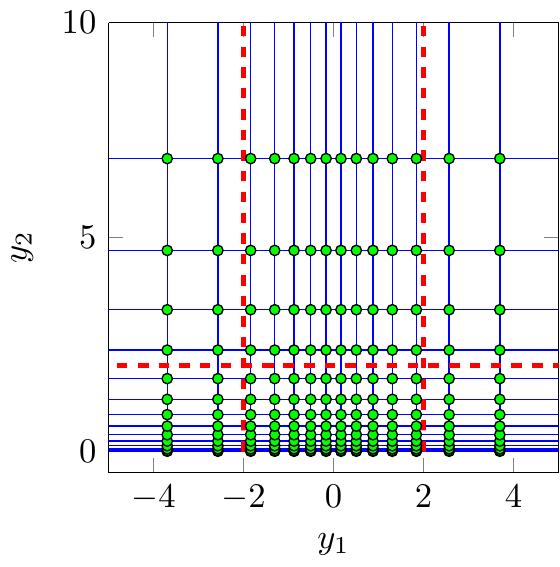}
		\caption{Gridlines used to discretise the density distribution $\nDensity$ in the half space.}
		\end{subfigure}\quad
		\begin{subfigure}[t]{0.47\textwidth}
		\includegraphics{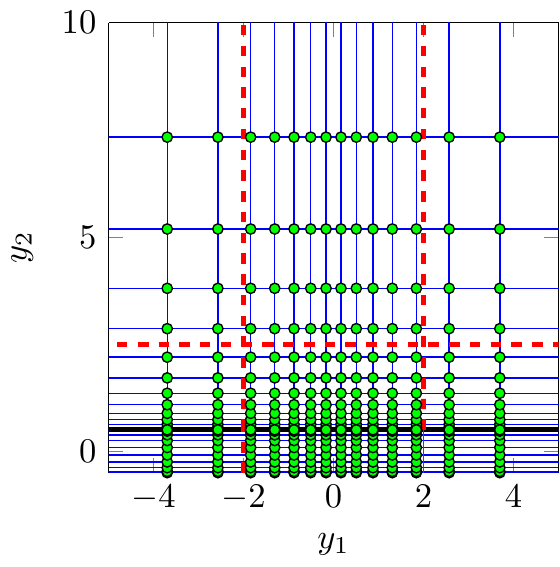}
		\caption{Gridlines used to discretise the half-space composed by a strip of height unity and a half-space.}
		\end{subfigure}
	\caption{\label{fig:HalfSpaceGrid} Gridlines of the discretisation of density $\nDensity$ and weighted densities $\nDensity_\alpha$. In the subplots above, $20$ collocation points are employed in each direction with mapping parameters $L_1 = L_2 = 2$. $50\%$ of the collocation points in $y_1$ and $y_2$-direction are located between the two vertical red dashed lines, defined by $\CartMap\klamm{\PhysSpace\klamm{\pm \frac{1}{\sqrt{2}},x_2}}$, and between the bottom of the half-space and the horizontal red dashed line, defined by $\CartMap\klamm{\PhysSpace(x_1,-1}$ and $\CartMap\klamm{\PhysSpace(x_1,0}$, respectively.}
\end{figure}

In order to correctly describe this behavior, we split the half-space
$\{(y_1,y_2):y_2> -1/2\}$ used to describe the weighted densities, into the
infinite slit $\{(y_1,y_2):y_2 \in [-1/2,1/2]\}$ and the half-space
$\{(y_1,y_2):y_2>1/2\}$, which are then discretised separately with
pseudospectral methods. In particular, the slit is described by
\begin{align}
\PhysSpace_{\Slit}(x_1,x_2) = \klamm{ \AlgMapWhole(x_1),\frac{x_2}{2}}, \label{eq:SlitDiscretization}
\end{align}
and the half-space is described by Eq.~\refe{eq:HalfSpaceDiscretization}, but shifted by $1/2$ in the positive $y_2$-direction, such as shown in Fig.~\ref{fig:HalfSpaceGrid}.
In the computations which we present here, the number of collocation points in this strip, $N_{\text{strip}}$, is chosen empirically to be the next even integer to $N_2/3$.  
This value was found to provide an accurate representation of the density and the weighted densities in the vicinity of the wall. 

\begin{figure}
	\centering
	\begin{subfigure}[t]{0.47\textwidth}
	\includegraphics{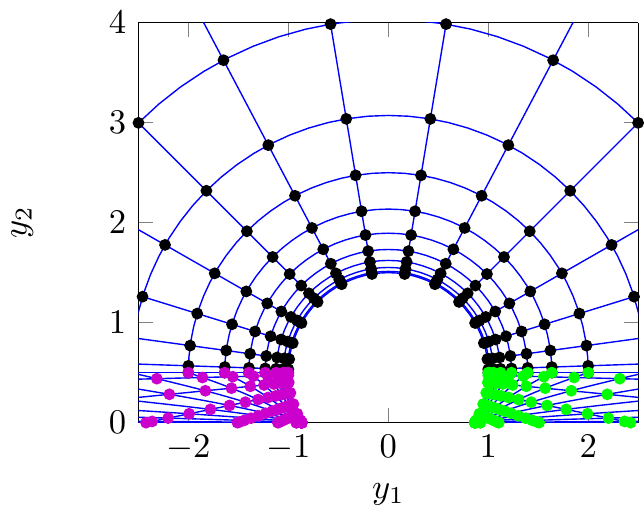}
	\caption{$\{{\bf r}: 1 \leq|\pos - \klamm{0,0.5}^T|\} \cap \{\pos:y_2 > 0\}$}
	\end{subfigure}\quad
	\begin{subfigure}[t]{0.47\textwidth}		
		\includegraphics{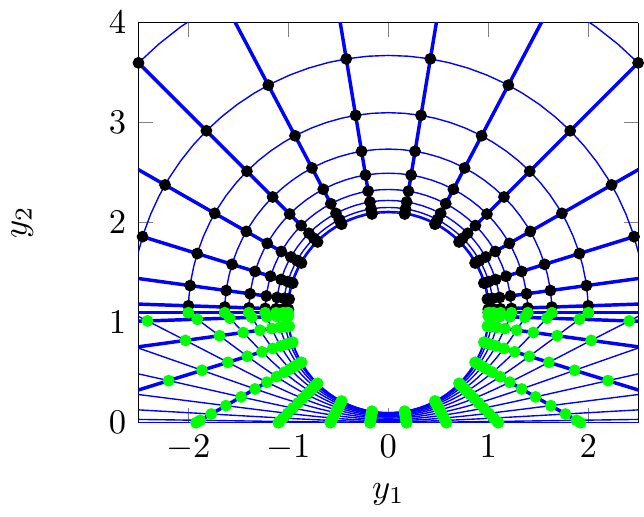}
	\caption{$\{{\bf r}: 1 \leq|\pos - \klamm{0,1.1}^T| \} \cap \{\pos:y_2 > 0\}$}
	\end{subfigure}
	\begin{subfigure}[t]{0.47\textwidth}
		\includegraphics{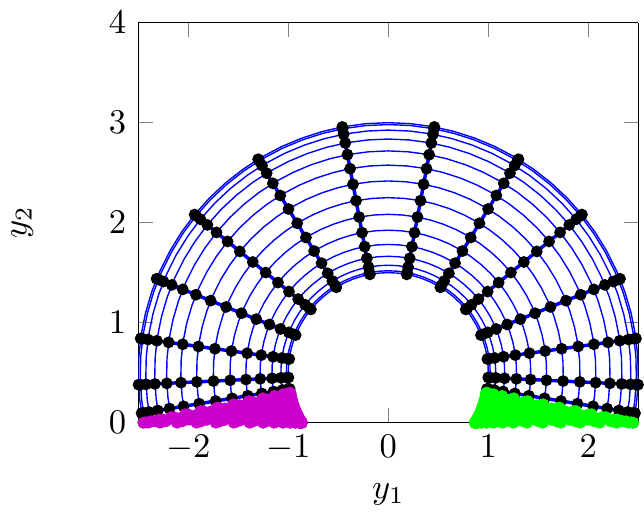}
		\caption{$\{{\bf r}: 1 \leq|\pos - \klamm{0,0.5}^T| \leq 2.5\} \cap \{\pos:y_2 > 0\}$}
	\end{subfigure}\quad
	\begin{subfigure}[t]{0.47\textwidth}		
		\includegraphics{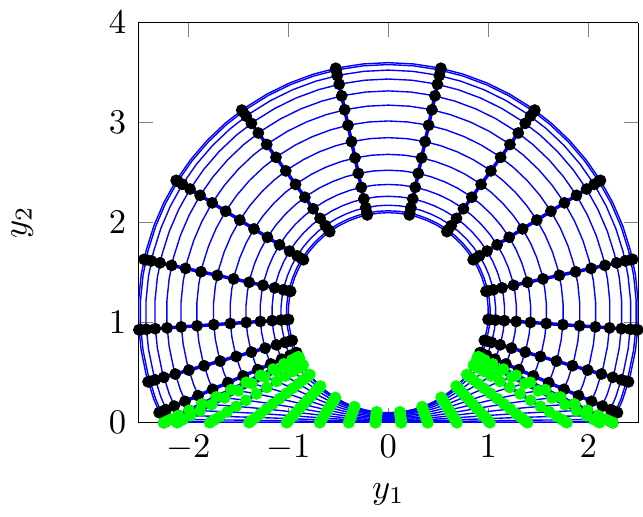}
		\caption{$\{{\bf r}: 1 \leq|\pos - \klamm{0,1.1}^T| \leq 2.5\} \cap \{\pos:y_2 > 0\}$}
	\end{subfigure}	
	\caption{\label{fig:IntersectionsExamples} Discretisation of intersections $\Intersection$ between the support of $\PhiTwoD^{\text{attr}}$ (see Eqs. (\ref{eq:phiattr:2D:1}), (\ref{eq:phiattr:2D:2})) with and without cutoff at $r_c = 2.5$ and the half-space, as given in Table \ref{tab:AssemblingShapesForIntersections} (see Sec.~\ref{Discretization} of the Appendix). Collocation points of different colors represent geometries which are discretised separately.}
\end{figure}

In order to compute the convolutions as described in Sec.~\ref{S:convolution}, one also has to discretise the intersections between the support of the weight function and the support of the density distribution. In Fig.~\ref{fig:IntersectionsExamples}, examples for such discretisations are given for intersections of the support of attractive particle--particle interactions and the half-space. A detailed description of the discretization of all geometries needed for the computation of the DFT-FMT model equations in the half-space is given in Sec.~\ref{sec:App:RepPhys}.

The methods described here can also be used to describe a slit domain between two parallel walls. Similarly, one could generalize the method to a domain bounded by a nonplanar, but smooth surface, by adapting the maps from the computational to the physical domain. However, for more complex geometries, e.g. involving corners, the representation of the weighted densities and the representation of the overlap between the sphere and the wall pose considerable challenges. In this case, one way forward is to develop a spectral element-like approach, where the domain is split into subdomains. Local complexities of the geometry, where the weighted densities are known not to be smooth, are then accounted for by robust and simple discretization schemes, whilst the large part of the domain in which the density and the weighted densities are smooth, are still spectrally represented. Alternatively, the domain may be split into smaller subdomains, so that the solution is represented spectrally in each.

In general, the concepts described here can also be applied to the 1D case.
In particular, the convolutions with weight functions with finite support
will reduce to integrations over intervals, as opposed to the more
complicated discretisations appearing in the 2D setting. Here, however, all computations for the cases exhibiting invariance in
$y_1$-direction are performed by formally discretising the first computational variable
$x_1$ with one collocation point only. This enables us to use the same
numerical scheme as in 2D.

\subsection{Iterative scheme}

We have implemented both
Newton~\cite{FrinkSalinger:2000kx,DouglasFrink:2000425} and
Picard~\cite{Roth:2010fk} iterative schemes to solve the Euler-Lagrange
equation~(\ref{eq:EulerLagrangeEquation}). We note that if the density of the
system is very small, then the ideal-gas contribution to
(\ref{eq:EulerLagrangeEquation}) diverges logarithmically. For example, small
densities can occur if the external potential $\Vext$ diverges. Therefore, in
order to avoid a special treatment for the density profile at zero, we solve
instead for
\begin{align}
z \defi \beta^{-1} \log \nDensity + {\bar V}_{\text{ext}}.
\end{align}
Here, the external potential was split into a bounded time-dependent part $\Vext'$ and a remainder ${\bar V}_{\text{ext}}$
\begin{align}
\Vext\klamm{y_1,y_2,t} = \Vext'\klamm{y_1,y_2,t} + {\bar V}_{\text{ext}}\klamm{y_1,y_2} .
\end{align}
In other words, ${\bar V}_{\text{ext}}$ is the time-independent unbounded
contribution to the external potential, such that $\Vext'$ is bounded. If
$\Vext$ is bounded, then ${\bar V}_{\text{ext}}$ is zero. We implemented the
simple iterative scheme
\begin{align}
z_{n+1} = z_n + \relaxationPar \Delta z,
\end{align}
where $\relaxationPar \in (0,1]$ is a relaxation parameter and the step size is defined by
\begin{align}
{\bf J} \klamm{\Delta z}_{\text{Newton}} &= - \frac{\delta \Omega[\nDensity]}{\delta \nDensity({\pos})}
\qquad \text{and} \qquad
\klamm{\Delta z}_{\text{Picard}} = - \frac{\delta \Omega[\nDensity]}{\delta \nDensity({\pos})} \label{eq:IterativeSteps}.
\end{align}
Here, ${\bf J}$ is the Jacobian of $\frac{\delta \Omega[\nDensity]}{\delta
\nDensity({\pos})}$ with respect to $z$. When solving for a DFT-FMT system
with the Newton scheme, we include the weighted densities
$\nDensity_2,\nDensity_3$ and ${\bf n}_2$ as unknowns, similar to the explicit
treatment of the nonlocal densities
in~\cite{FrinkSalinger:2002:NumericalChallenges}. We note that the
Picard iterative scheme makes use of the functional form of $\frac{\delta
\Omega[\nDensity]}{\delta \nDensity({\pos})}$, which can be written as a
fixed-point equation for $z$.

For the Newton iterations, we perform $O(10)$ steps with a relaxation
parameter $\relaxationPar = 0.5$, and then set $\relaxationPar = 1$. For the
Picard iterations, we initially set $\relaxationPar = 0.01$, and subsequently
increase $\relaxationPar$ to $0.2$. A Picard iterative step has a lower
computational cost than one Newton iteration, given that in each
Newton iteration, a linear system of equations has to be solved [see
Eq.~(\ref{eq:IterativeSteps})]. This can be substantially improved by
employing preconditioners~\cite{heroux:2007vn}. The general idea is that the
increased computational cost of each Newton iteration is compensated for by
the smaller number of iterations needed compared to Picard iterations.

In 1D computations of a fluid in contact with a hard wall, the Newton scheme
converges in $O(30)$ iterations, while the Picard scheme converges in
$O(500-1000)$ iterative steps, leading to a comparable computational cost. In
2D computations of a contact line in contact with a wall, we find that the
Picard iterative scheme is significantly more efficient than
Newton iterations.
\section{\revision{Verification and validation} \label{sec:Validation}}

The accuracy of the numerical scheme is validated in the following steps.
First, we confirm that the procedure of Sec.~\ref{S:convolution} to
compute convolutions does indeed converge. We then analyse the behavior of a
hard-sphere fluid with and without long-range interactions in contact with a
hard wall and confirm the accuracy of the results by checking whether the
so-called thermodynamic sum rules apply. Similarly, in 2D, we present
numerical results for a contact line and confirm the accuracy of the
computations by checking the force balance normal to the interface. We then
compare DFT results for multiple species of particles to the corresponding
particle distributions computed by slice sampling (a Markov chain MC
algorithm). In a final step, we perform DDFT computations of some model
systems and check for the mass conservation of the systems.

\subsection{Convergence test for convolutions}
\begin{figure}[ht]
	\centering
	\includegraphics{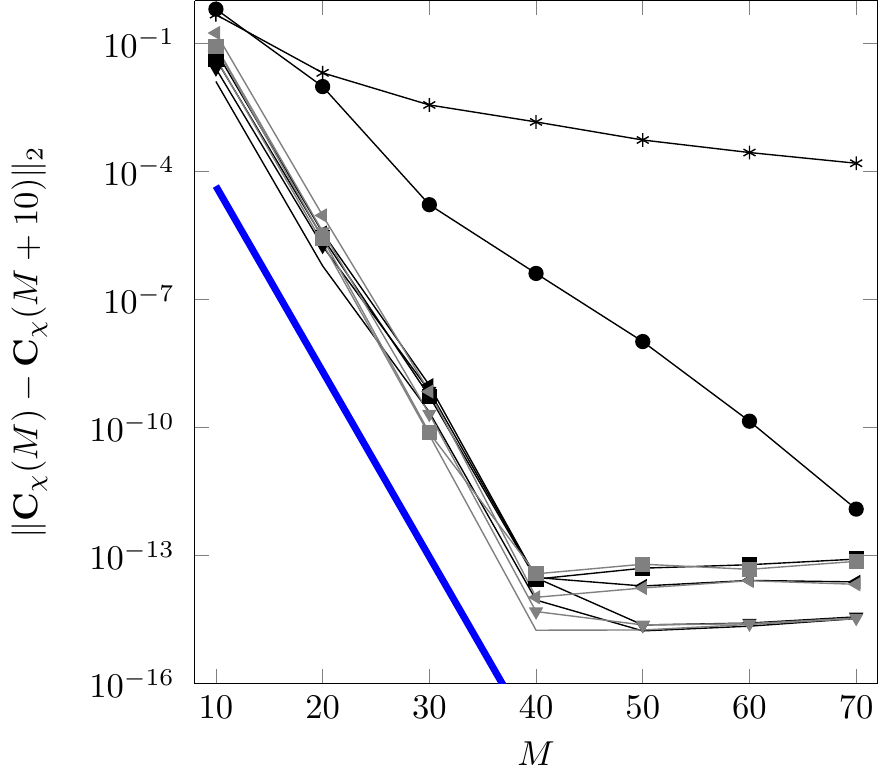}	
	\caption[Validation: Convergence of the convolution operator $\ConvOp$]{ \label{fig:ConvolutionMatrixError}
	Convergence of convolution matrices $\ConvOp$, as the number of collocation points $M$ 
	used to discretise the intersections $\Intersection$ is increased. For details on how 
	$\ConvOp$ is computed, see Sec.~\ref{S:convolution}. Results are given for the particle--particle 
	interaction potential $\BHattr$ without cutoff ($\ast$) and with cutoff radius $r_c = 2.5$ ($\bullet$) 
	(see Eq.~\refe{eq:pattr}), as well as for FMT weights $\weight_{2}$ ($\blacksquare$), $\weight_{3}$ ($\blacklozenge$), $\klamm{\boldsymbol \weight_{2}}_1$ ($\blacktriangledown$) and $\klamm{\boldsymbol \weight_{2}}_2$ ($\blacktriangleleft$) (see Eqs.~\refe{eq:WeightedDensitiesRFMT2}). Black and grey symbols denote the error of convolution matrices acting on functions defined on the half-spaces $\{(y_1,y_2):y_2 >0\}$
 and $\{(y_1,y_2):y_2 >-0.5\}$, respectively. The convergence is tested on the half-space introduced in Fig.~\ref{fig:HalfSpaceGrid} with $20 \times 20$ collocation points. \revision{As a guide to the eye, the blue solid line represents a steady convergence as $\sim e^{-M}$.}}
\end{figure}

When solving the Euler-Langrange equation (\ref{eq:EulerLagrangeEquation})
with an FMT model for the hard-sphere free energy contribution and an
attractive particle--particle free energy, convolutions have to be performed
in various steps. In Fig.~\ref{fig:ConvolutionMatrixError}, the convergence
of the convolution matrices $\ConvOp$ is tested. This is done by
comparing the incremental change of $\ConvOp$ as the number of collocation points $M$ (of each of the computational variables used to discretise the domain
$\Intersection\klamm{\pos}$), is increased.

Let us first distinguish between the convolutions which need to be computed
for the FMT part of the free energy and the convolutions for the attractive
contributions, the main difference being that the support of the FMT weight
functions $\weight_\alpha$ as defined in (\ref{eq:WeightedDensitiesRFMT2}) is
always finite and corresponds in 2D to a disk of diameter unity. In contrast,
the support of attractive particle--particle interactions \refe{eq:pattr},
which in 2D is given in Eqs.~\refe{eq:phiattr:2D:1}, \refe{eq:phiattr:2D:2}, is a larger disk with the
radius corresponding to the cutoff radius $r_c$, or, in the limiting case
of $r_c = \infty$, a support corresponding to the full 2D space. In
Fig.~\ref{fig:ConvolutionMatrixError}, we compare the convergence for a
weight function $\BHattr$ with cutoff radius $r_c = 2.5$, such as used e.g.
in~\cite{Parry:2014:WetGroove,Malijevsky:2013:CriticalPointWedgeFilling}, and
the full long-range particle--particle potential by setting $r_c = \infty$. As
expected, the convergence for $\BHattr(r_c=\infty)$ is not exponential. This
is because the support of the weight function is unbounded, and no
exponential convergence can be expected for unbounded domains other than for
special cases such as exponentially decaying weight functions. In contrast,
the convergence for a cutoff radius $r_c = 2.5$ is exponential.

For the FMT part of the free energy, the weight functions $\weight_\alpha$
are first convolved with the density distribution $\nDensity$ to compute the
weighted densities \refe{eq:nAlphaDef}. Then, the weight functions are
convolved with functions of the weighted densities given in
Eq.~\refe{eq:Rosenfeld_Variational} to compute the variation of the free
energy. The main difference between the two cases is that the densities and
the weighted densities are discretised using different domains, as shown in
Fig.~\ref{fig:HalfSpaceGrid}, and therefore the convolutions have to be
computed using different intersections $\Intersection$. As demonstrated in
Fig.~\ref{fig:ConvolutionMatrixError}, the convergence is exponential for all
cases.

We note that the convergence rates with respect to the number of collocation points $M$, depicted in Fig.~\ref{fig:ConvolutionMatrixError}, decrease with increasing number of collocation points $N$ of the main grid. This is because a higher number of modes of the density profile needs to be captured, therefore also requiring a higher value for $M$. 
\revision{In the following, we employ $M = 20 + N/4$ and $M = 20 + N/2$ collocation points for the computation of the convolution with the FMT weights $\weight_\alpha$ and the attractive potential with cutoff $r_c = 2.5$, respectively. The prefactors $1/2$ and $1/4$ account for the different convergence rates shown in Fig.~\ref{fig:ConvolutionMatrixError}.}

\subsection{Planar films and DFT sum rules}

\begin{figure}
	\centering
	\includegraphics{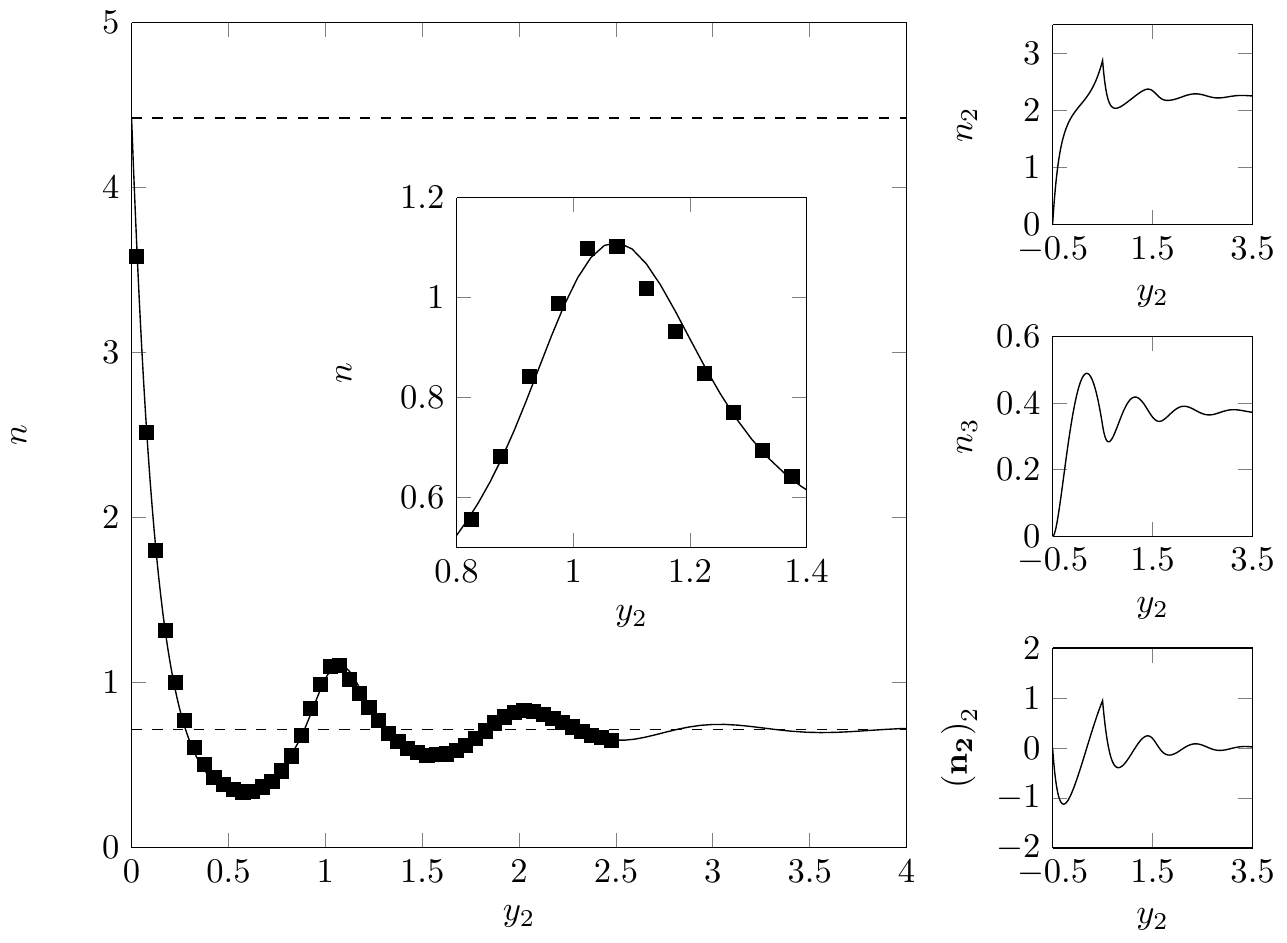}	
	\caption[Density profiles of a hard-sphere fluid in contact with a hard wall]{\label{fig:Density_DFTFMT} 
	The left subplot depicts the density profile of a hard-sphere fluid in contact with a hard wall, for bulk density $\nDensity_{\text{bulk}} = 0.7151$. The solid line represents results of a \revision{DFT-FMT} computation and the top and bottom dashed lines represent the values of the contact and bulk densities, respectively. Squares are MC computations by Groot \etal~\cite{Groot:1987uq}. The right subplots represent the weighted densities $\nDensity_2$, $\nDensity_3$ and the second component of the vector-weighted density ${\bf \nDensity}_2$ defined in Eq.~\refe{eq:nAlphaDef}. We employ $60$ collocation points in the direction normal to the wall.}
\end{figure}

We now proceed to compute planar wall-fluid interfaces for both pure hard-sphere fluids and hard-sphere fluids with attractive particle--particle interactions.
Due to the convergence properties shown in Fig.~\ref{fig:ConvolutionMatrixError}, we choose for the latter case a Barker-Henderson potential $\BHattr$ with radial cutoff $r_c = 2.5$.
We should note, however, that this effectively removes the long-range nature of the attractive particle--particle interactions.

In Fig.~\ref{fig:Density_DFTFMT}, results for a hard-sphere fluid in contact with a hard wall are compared with MC computations, giving a very good agreement. In the right subplots of Fig.~\ref{fig:Density_DFTFMT}, results for the weighted densities are depicted. It can be seen how the support of the weighted densities extends beyond the support of the density $\nDensity$, to $\{y_2 \geq -0.5\}$. Also, the discontinuity of the weighted densities at $y_2 = 0.5$ can be observed.

\begin{figure}
	\centering
	\begin{subfigure}[t]{0.47\textwidth}
	\includegraphics{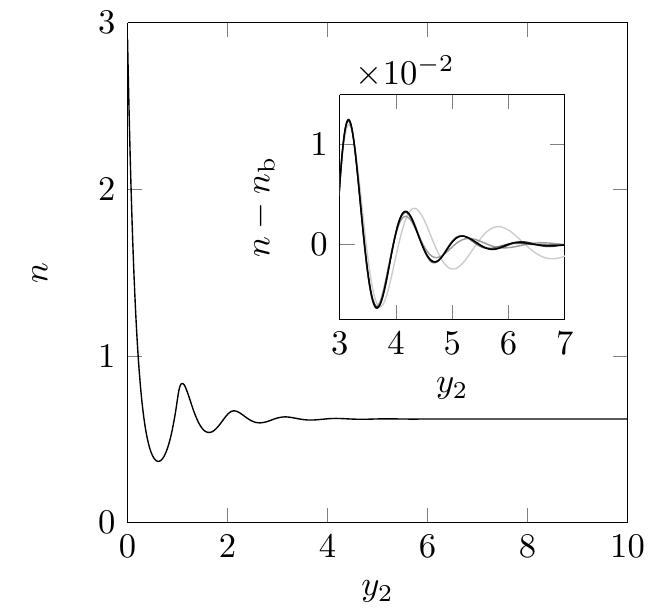}
	\caption{Hard-sphere fluid with bulk density $\nDensityBulk = 0.6220$ in contact with a hard wall.}
	\end{subfigure}
	\quad
	\begin{subfigure}[t]{0.47\textwidth}
	\includegraphics{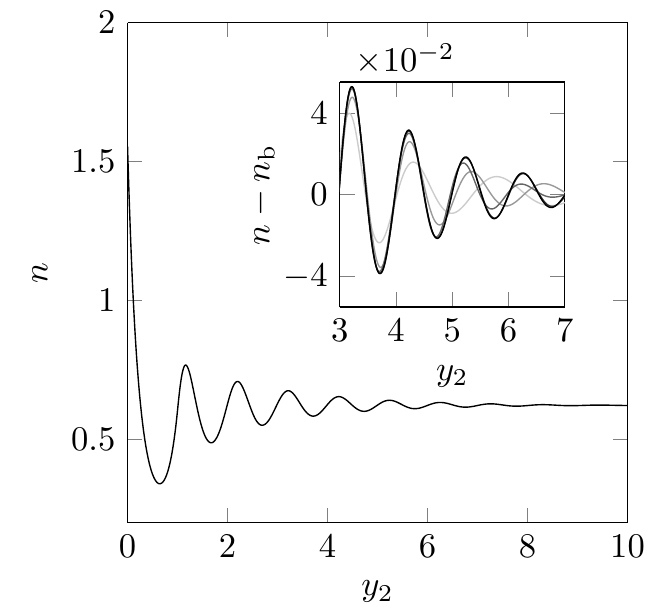}
	\caption{Barker-Henderson fluid with a cutoff radius $r_c = 2.5$ and an attractive wall (see Eq.~(\ref{eq:Vext:BH})) with parameter $\LJWdepth = 0.865$ at $k_B T = 0.75 \varepsilon$.}
	\end{subfigure}
	\caption[Validation: Density profiles at the wall-fluid interface]{\label{fig:DensityProfiles_Convergence} 
	Density profiles of fluids in contact with a wall. The insets are density oscillations around the bulk density $\nDensityBulk$ as the number of discretisation points is increased from \revision{50 to 150 in steps of 25}, where darker shades of grey represent a higher number of discretisation points. \revision{The darkest lines representing the highest numbers of collocation points are indistinguishable.}}
\end{figure}

For fluids with attractive particle--particle interactions, we employ a wall
with an algebraically decaying 9-3 potential
(e.g.~\cite{Nold:FluidStructure:2014,nold2015nanoscale})
\begin{align}
\Vext^{\text{BH}}(y_2) &=
4 \pi \depthLJW \klamm{ \frac{1}{45}  \klamm{\frac{1}{y_2+1}}^9 - \frac{1}{6}  \klamm{\frac{1}{y_2+1}}^3}, \label{eq:Vext:BH}
\end{align}
which is obtained by integrating an LJ potential over the
half-space $\{(y_1,y_2,y_3)\in \mathbb{R}^3| y_2 < 1\}$ and where $\depthLJW$
is the strength of the wall-fluid interaction potential. This parameter is
chosen such that the wall-liquid and the wall-vapor surface tensions are
equivalent (with the surface tensions defined as the excess grand potential
per unit area of the interface under consideration). All computations for
simple fluids are performed at saturation chemical potential, such that bulk
liquid and bulk vapor pressures are the same and hence both phases are
equally stable. Hard-sphere fluids without attractive particle--particle
interactions on the other hand are uniquely defined by their bulk densities
and do not allow for two equally stable phases. Such fluids are always
computed in contact with a hard, non-attractive wall. These restrictions are
employed to reduce the number of pertinent physical parameters and to ensure
comparability of the results.

The mapping parameter $L_2$ for the algebraic map
(\ref{eq:SemiInfiniteAlgMapping}) is chosen through a sensitivity study. In
particular, for values $L_2 \in [0.5,3.5]$ and for $N_2 = 100$ collocation
points, the sum rule error is in the range of $5 \pm 4 \times 10^{-5}$ for
the case of a fluid with attractions as well as for a hard-sphere fluid. This
suggests a low sensitivity of the results with respect to the mapping
parameter for all but very large or very small values of $L_2$, and we
therefore choose $L_2 = 2$ for all further computations.

Let us now study the properties of a typical fluid density profile close to a
hard wall. While wall-fluid density profiles do not exhibit any oscillations
for very low bulk (or vapor) densities, the same is not true for higher bulk
densities. In this case, we observe packing at the wall such as also depicted
in Fig.~\ref{fig:Density_DFTFMT}. Figure~\ref{fig:DensityProfiles_Convergence}
illustrates the difference between the quickly decaying density oscillations
of a pure hard-sphere fluid in contact with a hard wall and the much slower
decay of the oscillations if interparticle attractions and a long-range
attractive wall potential are added to the free energy. In particular, the
insets of Fig.~\ref{fig:DensityProfiles_Convergence} show the convergence of
the density profiles with increasing number of collocation points.

\begin{figure}[ht]
	\centering
	\begin{subfigure}[t]{0.47\textwidth}
		\includegraphics{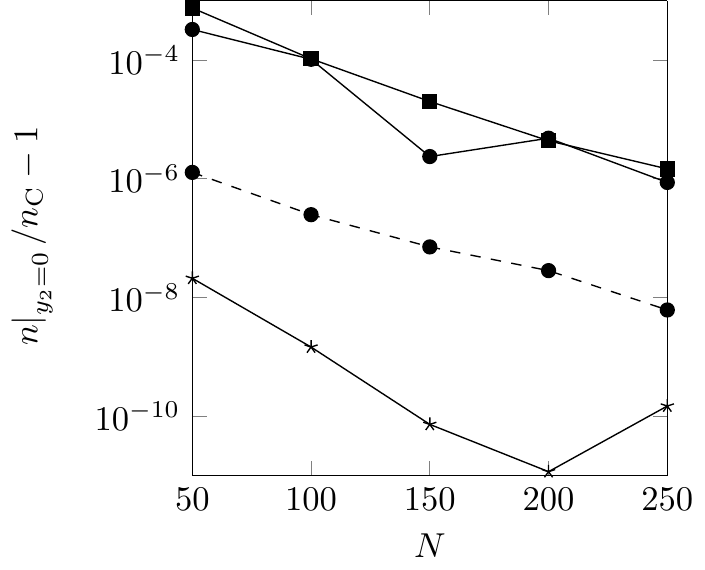} 
		\caption{Relative sum rule error of the contact density $\nDensityContact$.}
		\end{subfigure}		
		\quad
		\begin{subfigure}[t]{0.47\textwidth}
		\includegraphics{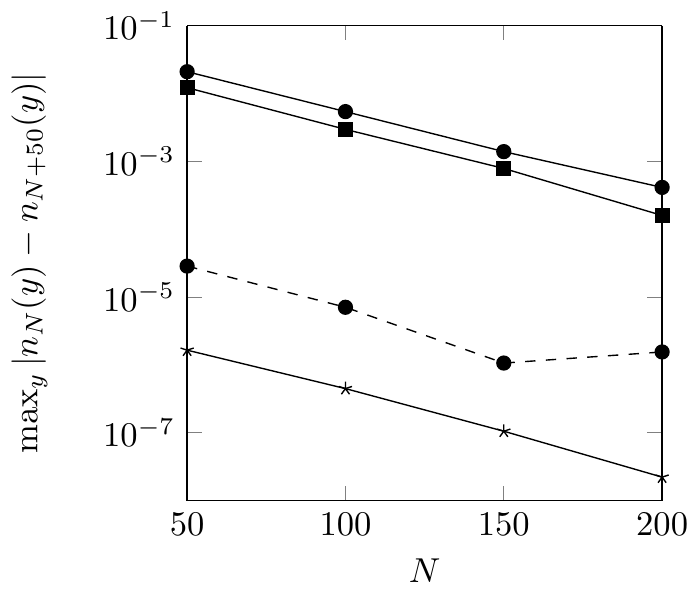} 
		\caption{Maximum incremental change of the density profile. \label{fig:SumRuleError:b}}
		\end{subfigure}
	\caption[Validation: Convergence test employing thermodynamic sum rules]{\label{fig:SumRuleError}
	Convergence of contact density and maximum incremental change of density profile for a fluid in contact with a wall as function of the number of grid points in direction normal to the wall.
	Computations are done for:
	(1) a hard-sphere fluid with attractive interactions ($\bullet$: $\BHattr\klamm{r_c = 2.5}$), in contact with an attractive wall (see Eq.~(\ref{eq:Vext:BH})), with $\LJWdepth = 0.865$ and $k_B T = 0.75 \varepsilon$. Wall-liquid and wall-vapor interface computations are denoted with solid and dashed lines, respectively; (2) a hard sphere fluid in contact with a hard wall, with bulk densities corresponding to the vapor and liquid densities of the Barker Henderson fluid, $0.028$ ($\ast$) and $0.622$ ($\blacksquare$), respectively.}
\end{figure}

Typically, the numerical accuracy of DFT computations is checked through sum
rules, which represent statistical mechanical connections between
thermodynamic properties of the system and the density
profile~\cite{Roth:2010fk}. The most prominent example is the so-called
contact theorem, which relates the density at a planar hard wall with the
bulk pressure of the system through
\begin{align}
  \left.\nDensity\right|_{y_2 = 0}  = \beta p + \beta \int_0^\infty \nDensity\klamm{y_2} \frac{\dI \Vext}{\dI y_2} \dI y_2 . \label{eq:ContactTheorem}
\end{align}
As an aside, we note here that the sign of the second term on the right hand side differs from the expression given in the review article by Roth~\cite{Roth:2010fk}, but is consistent with Refs.~\cite{SwolHenderson:1989,Herring:2010vn}. In Fig.~\ref{fig:SumRuleError}, we show the convergence of the relative sum rule error with the number of collocation points.

For the fluid with interparticle attractions, the error for the wall-vapor
interface is much lower than for the wall-liquid interface. This is due to
the absence of oscillations in the density profile at the wall-vapor
interface. Also, whilst the accuracy of the sum-rule for the wall-liquid
interface is lower, the relative sum rule error converges at a comparable
rate to that of a hard-sphere fluid of a similar bulk density. This result
suggests that the main factor limiting the convergence at the wall are
hard-sphere packing effects.

Whilst the sum rule provides an accurate check for consistency of the results
with thermodynamic principles, it is limited to the behavior of the density
close to the wall and therefore does not provide a convergence check for the
full density profile. We therefore show in the right subplot of
Fig.~\ref{fig:SumRuleError} the maximal incremental change of the density
profile as the number of collocation points is increased. Evidently, the
convergence is of a similar rate, but more smooth compared to the convergence
of the sum rule error. In particular, the results for the Barker-Henderson
wall-liquid interface converge slower than the results for a
hard-sphere fluid. This can be linked to the slower decay of the oscillations
for the Barker-Henderson fluid (see
Fig.~\ref{fig:DensityProfiles_Convergence}).

\subsection{The liquid-vapor contact line}
If a fluid with attractive particle--particle interactions is at saturation
chemical potential, a liquid can be in stable contact with a vapor, forming
an interface with a steep but smooth fluid density profile. When this
liquid-vapor interface is brought in contact with a substrate, and the
difference between the wall-liquid and the wall-vapor surface tensions is
not greater than the liquid-vapor surface tension, a well-defined
equilibrium contact angle is formed. This was also studied
in~\cite{Antonio2010} using the so-called local density approximation which
approximates the repulsive contribution to the free energy by employing
locally an equation of state (valid for the bulk fluid phase, but
nevertheless used for the inhomogeneous case in the presence of a wall). 
Unlike FMT, this does not capture the oscillatory structure of the fluid in the
vicinity of walls, which was later studied in Refs.~\cite{Nold:FluidStructure:2014,nold2015nanoscale} for an equilibrium contact line using the \revision{original FMT framework introduced by Rosenfeld}.

For the sake of simplicity, we focus on the case where the wall-vapor
surface tension $\surfaceTensionWV$ is equal to the wall-liquid surface tension
$\surfaceTensionWL$. In this case, the force balance parallel to the wall,
known as the Young-Laplace equation, implies that the contact angle $\theta$
between the liquid-vapor interface and the substrate is $90^\circ$:
\begin{align}
\surfaceTensionLV \cos \theta = \surfaceTensionWV - \surfaceTensionWL,
\end{align}
where $\surfaceTensionLV$ is the liquid-vapor surface tension and $\theta$
is the Young contact angle.

\begin{figure}
	\centering
	 \begin{subfigure}[t]{0.47\textwidth}
	 	\includegraphics{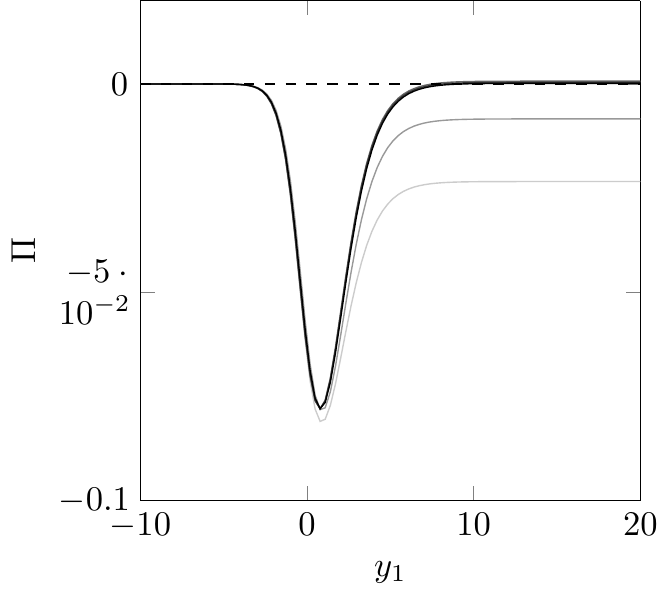} 
		\caption{\label{fig:SumRuleError2D:a}
		Disjoining pressure along the wall (see Eq.~(\ref{eq:DisjoiningPressureDefinition})). Darker shades of grey are results with a higher number of collocation points $N$.
		}
		\end{subfigure}
		\quad
	\begin{subfigure}[t]{0.47\textwidth}
	 	\includegraphics{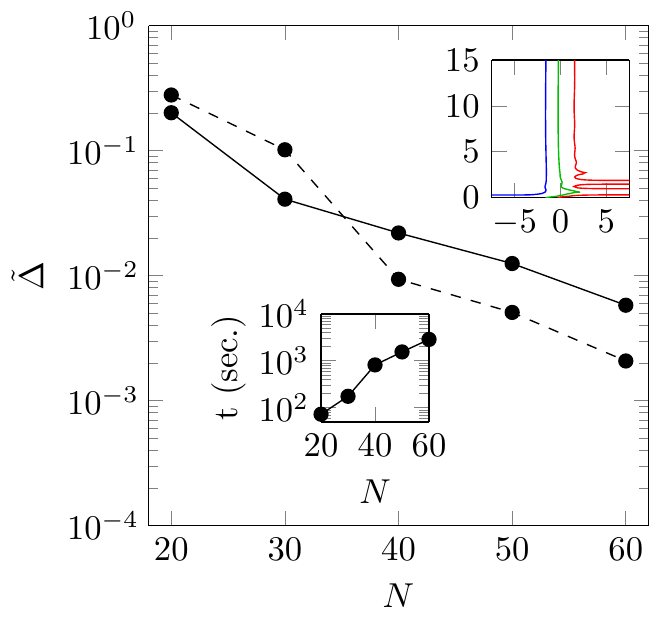} 
		\caption{\label{fig:SumRuleError2D:b}
		Relative error of the normal force balance (see Eq.~(\ref{eq:Numerics:defineDeltaDP}), solid lines), 
	and error in the planar limit, $\Delta \DisjoiningPressure$ (see Eq.~\refe{eq:Numerics:DeltaDP}, dashed lines) as a function of the number of collocation points $N$ in each direction.}		
		\end{subfigure}
	\caption[Validation: Normal force balance at the contact line.]{\label{fig:SumRuleError2D}
	Disjoining pressure and normal force balance at a $90^\circ$ contact angle.
	The top right inset of the right subplot shows a typical contour plot of the density profile, 
	with values $\nDensityV + \{0.05,0.5,0.95\}\klamm{\nDensityL-\nDensityV}$ represented by blue, green and red solid lines, respectively. The bottom left inset of the right subplot shows the 
	computation time of the Picard iterative scheme as a function of the number of collocation points.}
\end{figure}

In our 2D computations, the contact angle is computed as follows. First, we define the mapping parameter $L_1$ of the algebraic map (\ref{eq:WholeLineAlgebraicMapping}) used in (\ref{eq:HalfSpaceDiscretization}).
Knowing that the width of the liquid-vapor interface is about $4\LJdiam$ at
temperature $k_B T = 0.75 \LJdepth$, we make use of the fact that $50\%$ of
the collocation points are mapped onto the interval $[-L_1,L_1]$ by the
algebraic map (\ref{eq:WholeLineAlgebraicMapping}). As we also aim to resolve
properties such as the pressure of the fluid on the substrate, which vary at
slightly larger distances from the contact line, we set $L_1 = 4$ for all
further computations.

Furthermore, we enforce the density profile at distances $y_2 > \ytwomax$
from the substrate to be equivalent to an equilibrium liquid-vapor interface
at an angle $\theta$ with respect to the substrate. This also assumes the
validity of the Young-Laplace equation far from the substrate.
Computationally, one can verify the validity of this assumption by gradually
increasing $\ytwomax$ and observing the behavior of the liquid-vapor
interfaces for $y < \ytwomax$. In all further computations, we set $\ytwomax
= 25$.

In addition to the Young-Laplace equation, which represents the force
balance across the contact line in the direction parallel to the wall, we can
also verify the accuracy of our computations by checking the force balance
normal to the wall. In short, the normal pressure acting from a hard
attractive wall on a strip of fluid at position $y_1$ is also known as the
disjoining pressure $\DisjoiningPressure$, which is defined as
\begin{align}
\DisjoiningPressure\klamm{y_1} = - \int_{0}^\infty \nDensity\klamm{y_1,y_2} \frac{\dI \Vext}{\dI y_2} \dI y_2 + \beta^{-1} \nDensity(y_1,0)- p, \label{eq:DisjoiningPressureDefinition}
\end{align}
where $p$ is the bulk pressure. Insertion of
Eq.~(\ref{eq:DisjoiningPressureDefinition}) into
Eq.~(\ref{eq:ContactTheorem}) confirms that in the planar equilibrium case,
the excess pressure acting on the substrate always vanishes. When the film
height varies, however, the disjoining pressure does not vanish, and the
integrated excess pressure has to correspond to the pulling force of the
liquid-vapor interface in the direction normal to the substrate:
\begin{align}
- \int_{- \infty}^\infty \DisjoiningPressure\klamm{y_1} \dI y_1 = \surfaceTensionLV \sin \theta.\label{eq:NormalForceBalance}
\end{align}

In Fig.~\ref{fig:SumRuleError2D:a}, we present results of 2D contact line computations for the disjoining pressure along the wall. It can be observed how the disjoining pressure approaches constant values 
\begin{align}
\DisjoiningPressure_\pm \defi \lim_{y_1 \to \pm\infty} \DisjoiningPressure(y_1)
\end{align}
away from the contact line, as the density profile normal to the wall converges to a planar equilibrium wall-liquid and wall-vapor interface. 
In agreement with the physical predictions, $\DisjoiningPressure_{\pm}$ vanishes as the number of collocation points is increased. In particular, the convergence of
\begin{align}
\Delta \DisjoiningPressure \defi \max |\DisjoiningPressure_\pm| \label{eq:Numerics:DeltaDP}
\end{align}
 is shown with the dashed lines in Fig.~\ref{fig:SumRuleError2D:b}. 

We note, however, that the integration of the disjoining pressure in the normal force balance 
(\ref{eq:NormalForceBalance}) is only well-defined if $\DisjoiningPressure$ vanishes
for $y_1 \to \pm \infty$. Consequently, the numerical error of $\DisjoiningPressure_\pm$ --- regardless of its size --- has to be accounted for when computing the normal force balance.
Here, we do so by restricting the integration in the normal force balance to values of $y_1$ for which the absolute value of the disjoining pressure is greater than $2\Delta \DisjoiningPressure$: $\{y_1:|\DisjoiningPressure(y_1)|>2\Delta \DisjoiningPressure\}$. The relative numerical error of the normal force balance \refe{eq:NormalForceBalance} then becomes
\begin{align}
\tilde \Delta  \defi \left| 1 + \frac{ \int_{\{y_1:|\DisjoiningPressure(y_1)|>2\Delta \DisjoiningPressure\}} \DisjoiningPressure\klamm{y_1} \dI y_1 }{\surfaceTensionLV \sin \theta} \right|.
\label{eq:Numerics:defineDeltaDP}
\end{align}
This is depicted in Fig.~\ref{fig:SumRuleError2D:b}, where the Young contact angle is computed via the Young-Laplace equation. 

\subsection{Multiple species -- comparison to slice sampling}

Here we consider systems of a fixed number of particles of different species.
The equilibrium states are computed similarly to the single-fluid case, by
solving the Euler-Lagrange equation (\ref{eq:EulerLagrangeEquation}) for
each species $k$ separately:
\begin{align}
\frac{\delta \Omega[\{\nDensity_k\}]}{\delta \nDensity_k({\pos})} = 0.
\end{align}
The different species pairs $i, j$ may experience different interparticle
potentials $\Phi_{\text{2D}}^{(i,j)}$. Also the external potentials
$\Vext^{(k)}$ acting on species $k$ may differ. We demonstrate the
convergence of the method with respect to the number of collocation points,
and compare the results to those obtained by slice sampling.
Details on the DDFT formulation for multi-species fluids as well as the
physics of such fluids are given in~\cite{Goddard:2013multi}.

Slice sampling~\cite{Neal03} is a Markov chain MC method for sampling from a
statistical distribution, in this case from the equilibrium configuration of
the particles.  A particular advantage of this method is that the results are
invariant under scaling the distribution by a constant; hence one does not
need to know the normalisation constant (or partition function). In
particular, we consider two systems, soft particles (which repel each other
with a Gaussian interparticle potential) in a hard wall box, and hard disks
in an infinite planar geometry.

\begin{figure}[h]
	\centering
	\includegraphics[width=0.8\textwidth]{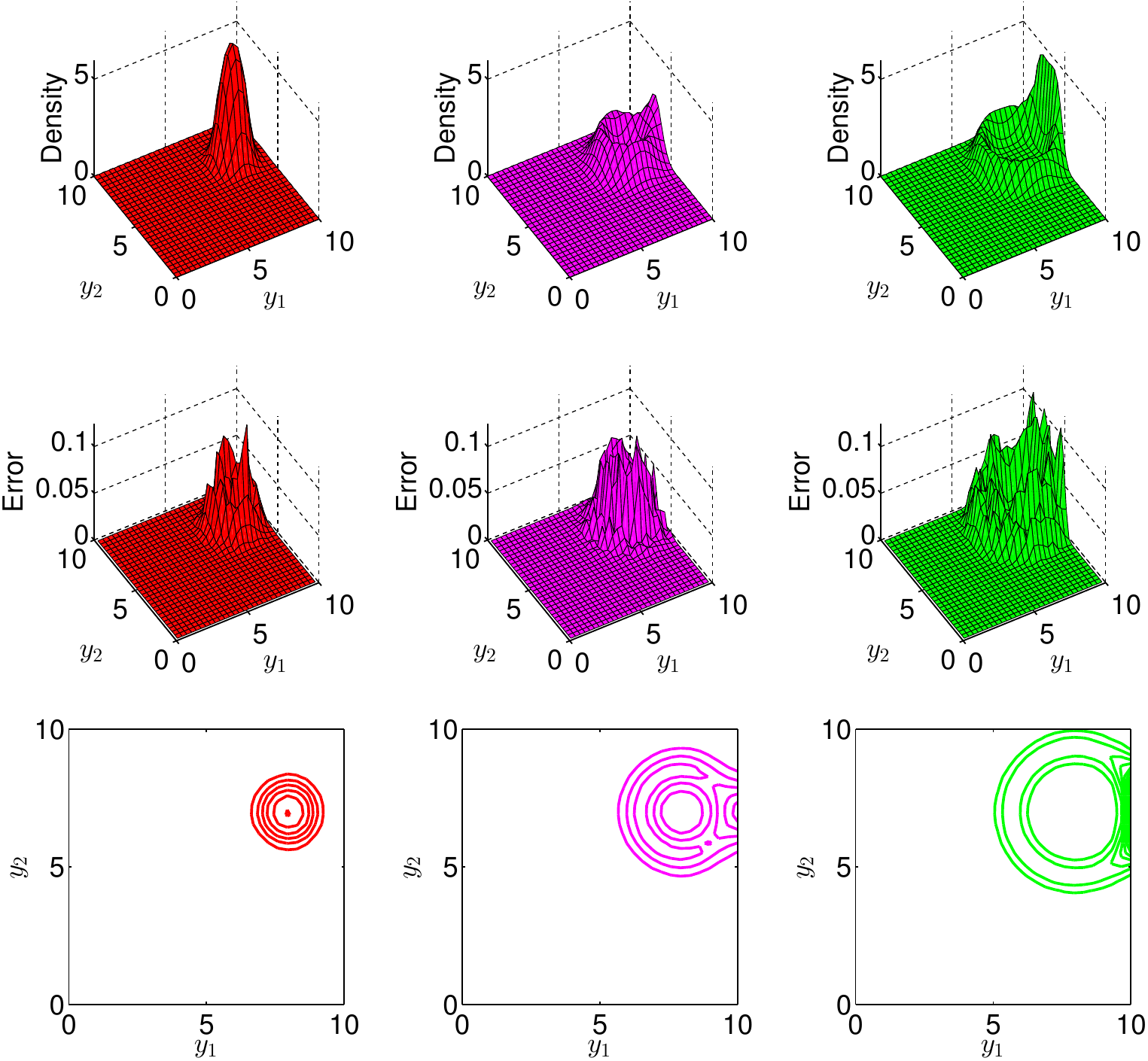}
	\caption{ \label{fig:multiSpeciesGaussian} Density profiles (top row), the absolute value of the difference between the DDFT result and that from stochastic sampling (second row) and contours (bottom row) for a system of three species interacting with Gaussian potentials (increasing size
	left--right) for 50 collocation points.}
\end{figure}

For the soft particles, the external potential is given by a quadratic
\[
	V_{\text{ext},\rm{S}}\klamm{y_1,y_2} = 2 \big( (y_1 - 8)^2 + (y_2 - 7)^2 \big)
\]
along with a confining hard-walled box of side length 10.  The system contains 
20 `small', 20 `medium' and 20 `large' particles, which represent different three different particle species. They repel each other via Gaussian potentials for each
pair of species $i$, $j$:
\[
	\Phi_{\text{2D},\rm{Gaussian}}^{(i,j)}(r) = 2 \exp \klamm{ -\frac{r^2}{\sigma_{i,j}^2}},
\]
where $r$ is the interparticle distance.  Here we define effective particle diameters $\alpha_1 = 0.5$, $\alpha_2 = 1$ and $\alpha_3 = 1.5$,
 which in turn define the length scale $\sigma_{i,j}$ of the interparticle potential via $\sigma_{i,j} = (\alpha_i + \alpha_j)/2$.
The density profiles for this system are given in Fig.~\ref{fig:multiSpeciesGaussian}, 
where it is also shown that the maximum relative error is about $2\%$ between DDFT and stochastic sampling for the parts of the density that have significant mass. Convergence with the number of collocation points is shown in Fig.~\ref{fig:App:multiSpeciesGaussianN} of Appendix~\ref{sec:App:Multispecies}.

\begin{figure}
	\includegraphics[width=\textwidth]{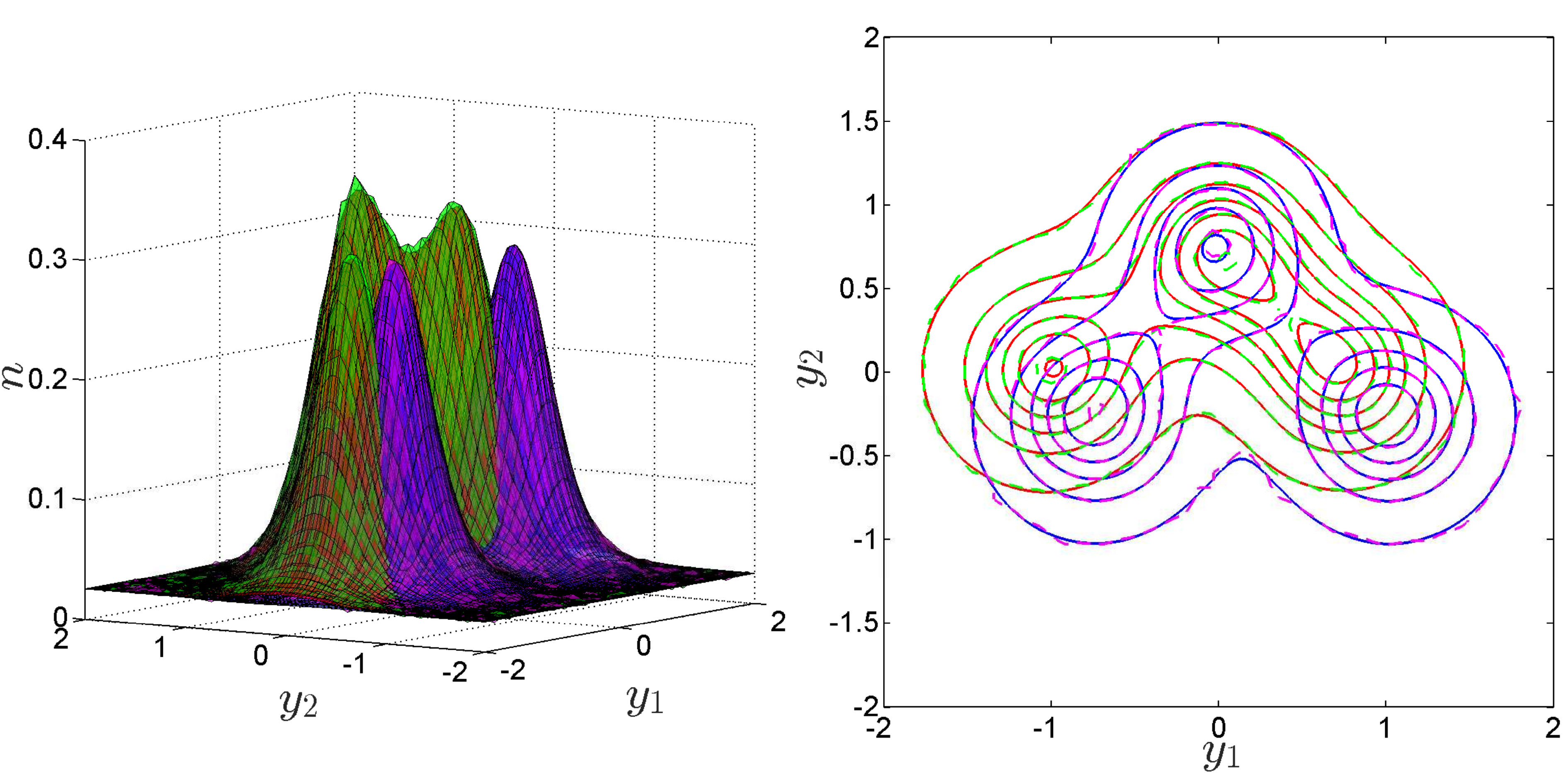}		
	\caption{ \label{fig:FMT2Species} 
	Comparison of density profiles (left subplot) and contours (right subplot) for a system of two hard disk species for 60 collocation points (solid lines: red, blue), virtually indistinguishable with results of slice sampling (dashed lines: green, magenta).}
\end{figure}

The external potential for the hard disk system consists of a weak confining potential $\bar V$ and
three Gaussian wells, which are in different positions
for the two species.
Here, we set the confining background potential to
\[
	{\bar V}_{\rm{HD}}\klamm{y_1,y_2} = 0.01 (y_1^2 + y_2^2)
\]
and employ attractive Gaussian wells, leading to the external potential for the $k$th species
\[
	\Vext^{(k)}\klamm{y_1,y_2} = {\bar V}_{\rm{HD}}  - 3 \sum_{\ell=1}^3
	e^{-2|{\bf y} - {\bf y}^{(k)}_\ell|^2}.
\]
Here, ${\bf y} = \klamm{y_1,y_2}$ and ${\bf y}^{(k)}_\ell$ represents the
position of the $\ell$-th external potential well for species $k$. For the
first species we set ${\bf y}^{(1)}_{\{1,2,3\}} =
\left\{\klamm{-1,0},\klamm{0.75,0},\klamm{0,0.75}\right\}$ and for the second
species ${\bf y}^{(2)}_{\{1,2,3\}} =
\left\{\klamm{-0.75,-0.25},\klamm{1,-0.25},\klamm{0,0.75}\right\}$.
The system contains 10 particles from each of the two species. The FMT
calculations are performed using 2D FMT for hard
disks~\cite{RothMeckeOettel2012} (see Sec.~\ref{sec:DiskFMT}). Computational
results for the density profiles are depicted in Fig.~\ref{fig:FMT2Species}, and 
convergence with the number of collocation points is shown in Fig.~\ref{fig:App:FMT2SpeciesN} of Appendix~\ref{sec:App:Multispecies}.

\revision{In Fig.~\ref{fig:MultispeciesConvergence}, we} consider the convergence of the method by computing the relative $L^2$
error of the density $\nDensity$ for pairs of computations with $N$ and $N+2$ collocation
points, with all other parameters fixed. This is done by interpolating each
result onto the same equispaced grid of $100 \times 100$ points and then
computing the $L^2$ norm of the difference,
$\|\nDensity_{N-2}-\nDensity_N\|_2$, and normalising by the norm of the
most accurate computation $\| \nDensity(N_{\max})\|$ (which, as shown in
Figs. \ref{fig:multiSpeciesGaussian} and \ref{fig:FMT2Species}, is virtually
indistinguishable from the slice sampled `exact' result). We plot the
maximum error over all species. For the box calculation, the equispaced grid
covers the whole box, whilst for the infinite domain it is restricted to
$[-2,2] \times [-2,2]$, outside of which the densities are very small. It is
clear that the method exhibits the expected spectral accuracy.

For the Gaussian case we use $M = 20 \times 20$ points to compute
the convolution matrices accounting for the mean field contributions. For the
FMT case we use $L=4$ in both directions for the infinite space, 10
collocation points for the circle FMT contributions and $10 \times 10$ points
for the disk contributions.  The results show very good agreement with those
obtained by doubling the number of points for the mean field and FMT
computations, and the computations show very little dependence on $L$ in the
region around the chosen values.

As mentioned, Figs.~\ref{fig:multiSpeciesGaussian} and \ref{fig:FMT2Species}
compare the DFT computations with the slice sampled data.  Slice sampling is
performed using MATLAB's `slicesample' routine, with 5,000,000 samples; the
resulting densities are then histogrammed into $50\times50$ boxes. Evidently
the agreement is very good. The slight deviations for higher densities in the
FMT computations are to be expected as FMT is less accurate in the
high-density regime.

\begin{figure}
\begin{center}
	\includegraphics[width=0.5\textwidth]{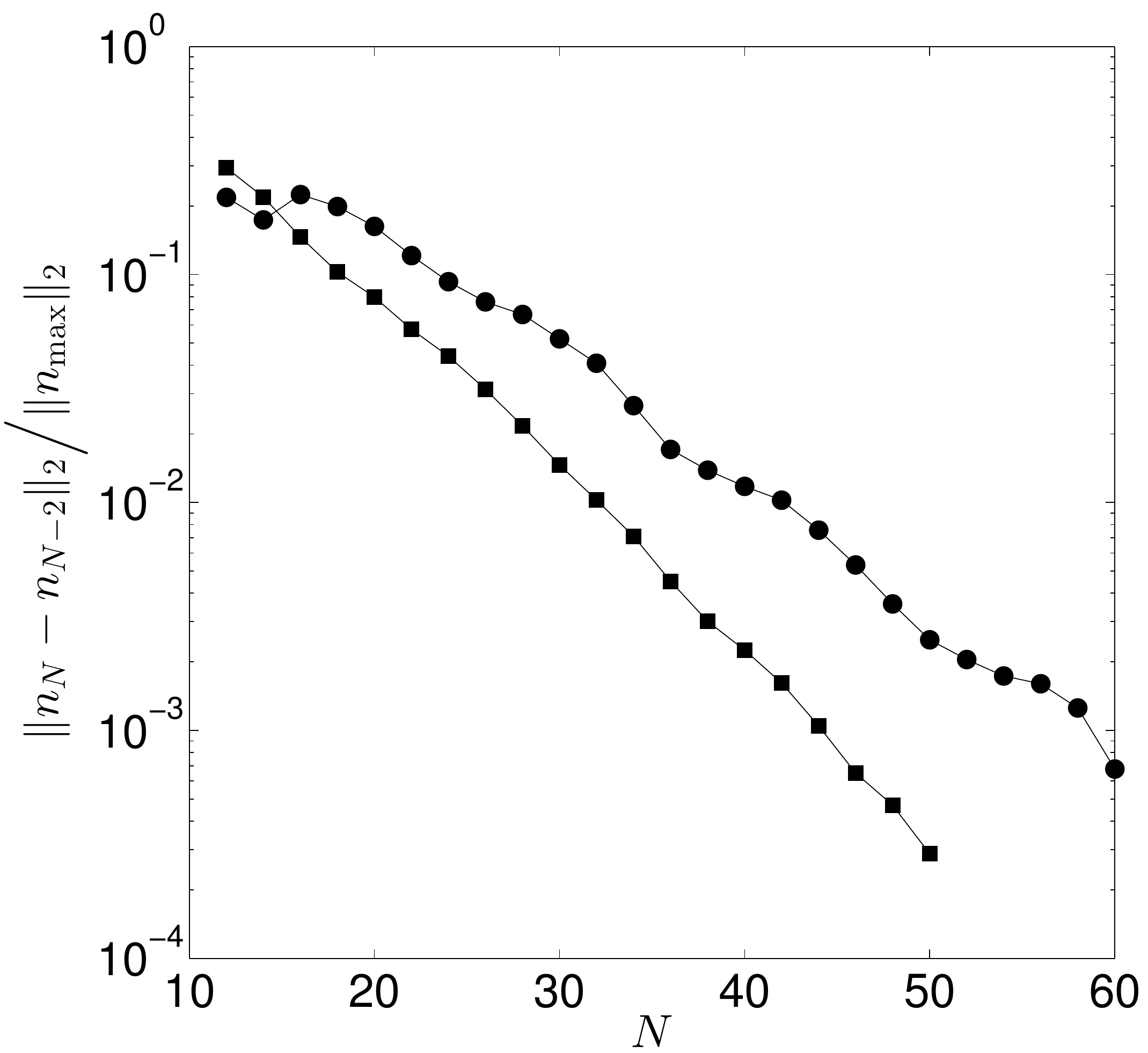}
\end{center}
	\caption{\label{fig:MultispeciesConvergence} Maximum relative $L^2$ errors of successive pairs
	of computations for $N$ and $N+2$ collocation points for soft particles ($\blacksquare$) and FMT hard
	disks ($\bullet$).}
\end{figure}

\subsection{Mass conservation of DDFT computations}
\begin{figure}
	\centering
	 	 \begin{subfigure}[t]{0.47\textwidth}
 		\includegraphics{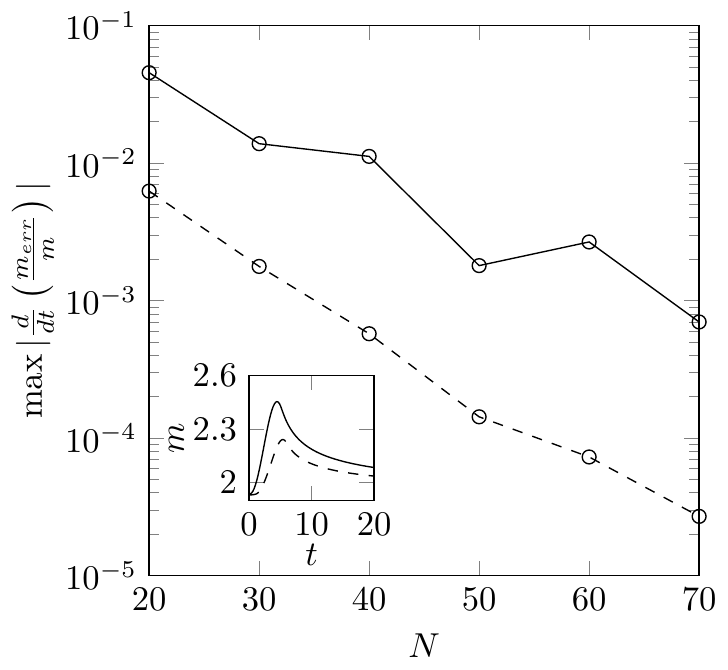}
		\end{subfigure}\quad
	 	 \begin{subfigure}[t]{0.47\textwidth}
		\includegraphics{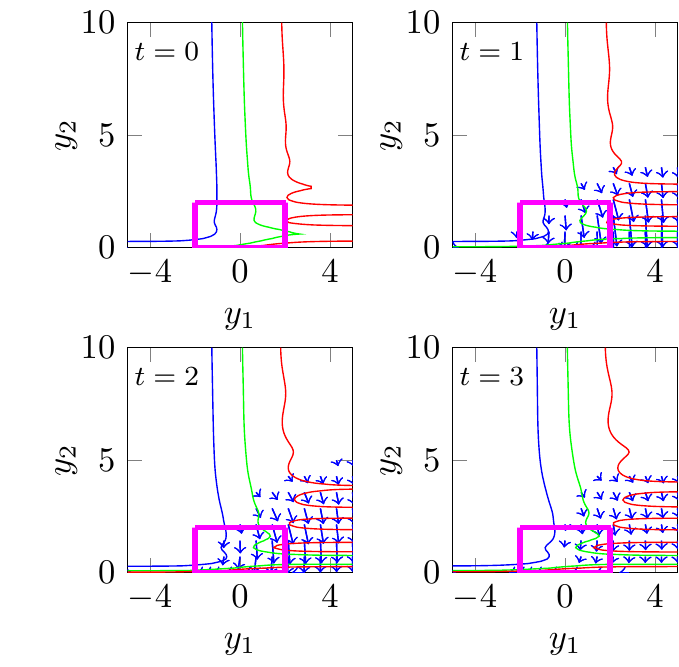}
		\end{subfigure}\quad
	\caption{\label{fig:DynamicMaxMassError} Relative mass error per unit time in the rectangular subdomain $[-2,2]\times [0,2]$.
	Left: the maximum relative mass error per unit time $\frac{\dI }{\dI t}(\merr/m)$ as a function of the number of collocation points, where the inset depicts the mass in the
    subdomain as a function of time. Solid lines show results for the
    overdamped limit with $\Delta \depthLJW = 1$,  $\tau = 5$ and dashed lines are the inertial case with friction
    coefficient $\gamma = 2$. Right: contour plots for
    four different times. \revision{The blue, green and red solid curves show the locus of points in the domain for which $\nDensity = \nDensityV + i\klamm{\nDensityL-\nDensityV}$, where $i = 0.05,0.5$ and $0.95$, respectively.} All computations are performed for a Barker-Henderson-FMT fluid modelled with a cutoff potential $\BHattr$ with dynamic external potential $V_1$ as defined in (\ref{eq:DynamicVext:1}).	
    The subdomain is discretized with $40 \times 40$ collocation points.}
	\end{figure}
	
\revision{In this section, we} demonstrate the effectiveness of our numerical scheme in tackling \revision{the DDFT equations introduced in Sec.~\ref{DDFT}. In particular,} we consider a simple scenario of the density changing between two different equilibrium configurations. The external potential is given by
\begin{align}
V_1(y,t;\depthLJW,\Delta\depthLJW,\tau) = \Vext^{\text{BH}}(y;\depthLJW)
\left\{
\begin{array}{ll}
1 + \frac{\Delta\depthLJW}{\depthLJW} \sin\klamm{\frac{\pi t}{\tau}}& \text{for } t < \tau\\
1& \text{for } t \geq \tau
\end{array}
\right. , \label{eq:DynamicVext:1}
\end{align}
where $\Vext^{\text{BH}}$ is defined in (\ref{eq:Vext:BH}).

Figure~\ref{fig:DynamicMaxMassError} shows the performance of our scheme as the
number of collocation points is increased, for both the overdamped and
inertial systems introduced in Sec.~\ref{DDFT}. \revision{We note that due to the unbounded nature of the geometry, the mass of the full system is infinite. In particular, local changes at the contact line lead to an influx of mass from the reservoir, thus increasing the net mass within a sub-domain near the contact line (see e.g. inset of left subplot of Fig.~\ref{fig:DynamicMaxMassError}). For any finite domain, this change in mass has to be accounted for by a mass influx through its boundaries. 
In the left subplot of Fig.~\ref{fig:DynamicMaxMassError}, this fundamental principle of mass conservation is used to validate the numerical scheme.}

The equations are solved using the
implicit variable-time-step solver for stiff problems `ode15s' from MATLAB. It is noteworthy that the inclusion of inertia triples the number of unknowns
in the problem as in addition to the density the two components of the
velocity must be evaluated.
\section{Conclusion \label{sec:Conclusion}}

Historically, the main focus of computational fluid dynamics has been on
analysing and finding numerical solutions to the NS equations and their
simplifications. At the nanoscale, however, these equations do not apply as
interparticle interactions, described by MD-MC, dominate. Unfortunately,
despite drastic improvements in computational power, MD-MC simulations are
only applicable for small fluid volumes. A compromise between the two is
classical DFT which retains the microscopic details of macroscopic systems
but at a cost significantly lower to that of macroscopic equations.
Typically, DFT relies on the formulation of a free energy functional which is
minimal at equilibrium, resulting in an integral equation for the (continuous)
fluid density. The integral terms are due to the modelling of van der Waals
attractive forces and hard-sphere effects via e.g.\ FMT~\cite{Wu-DFT}; the
advantage of FMT being that it captures the oscillatory behavior of the fluid
density in the vicinity of walls.

Typically, the corresponding DFT-FMT equations are solved using either
Fourier methods for the convolutions~\cite{Roth:2010fk}, or real-space
computations of the convolution grids~\cite{FrinkSalinger:2002:NumericalChallenges,TRAMONTO:2007}. These approaches,
however, formally restrict the applicability of the associated numerical
schemes to bounded domains, where Fourier methods also assume periodicity of
the solution. Here, we introduce an alternative numerical method based on a
pseudospectral scheme with the collocation points mapped onto unbounded domains~\cite{Boyd:2001Cheb}.

The resulting scheme is highly efficient, accurate and fast, and allows for
the resolution of density profiles and nanoscale fluid behavior with a
comparably lower number of collocation points. The convolutions required in
DFT-FMT computations are performed in real space, which allows for a
pre-computation of the convolution matrices. This process must be done only
once for every geometry and can easily be parallelized.

We apply our scheme to four different DFT scenarios. First, we compute the
density profile of a hard-sphere fluid with and without attractions close to
a hard wall, and demonstrate convergence to the so-called thermodynamic sum
rules~\cite{Roth:2010fk} with increasing number of collocation points. We
then apply the numerical scheme to the computation of an equilibrium contact
angle and, in addition to the Young-Laplace equation, i.e.\ the
tangential-stress balance at the contact line, we also confirm the
normal-force balance. Subsequently, the scheme is applied to multi-species
systems where it is successfully validated against Monte-Carlo computations.
Finally, we present dynamic computations for a contact line, by using the
DDFT framework developed in~\cite{Goddard:2013Unification} and we confirm the
accuracy of our scheme by testing for mass conservation.

We believe that the versatility of the outlined numerical
methodology allows for the treatment of more involved geometries, such as
capillaries and grooves, but also the inclusion of other nanoscale effects
such as hydrodynamic interactions~\cite{Goddard:2012general}, which are neglected in the
DDFT equations adopted here. Furthermore, of particular interest would be the
development of a multiscale algorithm to bridge the nanoscale, described here
by integral equations, with the macroscale described by NS. We shall address
these and related questions in future studies.
\section*{Acknowledgements}
We acknowledge financial support from Imperial College (IC) through a DTG
International Studentship \revision{and from the Engineering and Physical
Sciences Research Council (EPSRC) of the UK through Grants No. EP/L027186,
EP/L025159 and EP/L020564 as well as an EPSRC-IC Pathways to Impact-Impact
Acceleration Award, Grant No. EP/K503733.}

\bibliographystyle{plain}
\bibliography{Bibliography}

\begin{thebibliography}{10}

\bibitem{TRAMONTO:2007}
Tramonto 4.x.
\newblock \url{https://software.sandia.gov/DFTfluids/index.html}, 2007-present.

\bibitem{Archer.20090107}
A.~J. Archer.
\newblock Dynamical density functional theory for molecular and colloidal
  fluids: {A} microscopic approach to fluid mechanics.
\newblock {\em J. Chem. Phys.}, 130(1):014509--8, 2009.

\bibitem{Archer:2004mb}
A.~J. Archer and R.~Evans.
\newblock Dynamical density functional theory and its application to spinodal
  decomposition.
\newblock {\em J. Chem. Phys.}, 121(9):4246, 2004.

\bibitem{ArcherEvans:2013}
A.~J. Archer and R.~Evans.
\newblock Relationship between local molecular field theory and density
  functional theory for non-uniform liquids.
\newblock {\em J. Chem. Phys.}, 138(1):014502, 2013.

\bibitem{Berrut:2004ly}
J.-P. Berrut and L.~N. Trefethen.
\newblock Barycentric {L}agrange interpolation.
\newblock {\em SIAM Review}, 46(3):501, 2004.

\bibitem{Lasse:etal:2013}
L.~B{\o}hling, A.~A. Veldhorst, T.~S. Ingebrigtsen, N.~P. Bailey, J.~S. Hansen,
  S.~Toxvaerd, T.~B. Schrøder, and J.~C. Dyre.
\newblock Do the repulsive and attractive pair forces play separate roles for
  the physics of liquids?
\newblock {\em J. Phys.: Condens. Matter}, 25(3):032101, 2013.

\bibitem{Bonn.20090527}
D.~Bonn, J.~Eggers, J.~Indekeu, J.~Meunier, and E.~Rolley.
\newblock Wetting and spreading.
\newblock {\em Rev. Mod. Phys.}, 81:739--805, 2009.

\bibitem{Boyd:198243}
J.~P. Boyd.
\newblock The optimization of convergence for {C}hebyshev polynomial methods in
  an unbounded domain.
\newblock {\em J. Comput. Phys.}, 45(1):43, 1982.

\bibitem{Boyd:2001Cheb}
J.~P Boyd.
\newblock {\em Chebyshev and Fourier spectral methods}.
\newblock Dover publications, 2001.

\bibitem{Cao_Wu:2004}
D.~Cao and J.~Wu.
\newblock Density functional theory for semiflexible and cyclic polyatomic
  fluids.
\newblock {\em J. Chem. Phys.}, 121(9):4210, 2004.

\bibitem{Chan:2005uq}
G.~K.-L. Chan and R.~Finken.
\newblock Time-dependent density functional theory of classical fluids.
\newblock {\em Phys. Rev. Lett.}, 94(18):183001, 2005.

\bibitem{ChandlerMcCoy:1986}
D.~Chandler, J.~D. McCoy, and S.~J. Singer.
\newblock {Density functional theory of nonuniform polyatomic systems. I.
  General formulation}.
\newblock {\em J. Chem. Phys.}, 85(10):5971, 1986.

\bibitem{clenshaw:1960}
C.~W. Clenshaw and A.~R. Curtis.
\newblock A method for numerical integration on an automatic computer.
\newblock {\em Numer. Math.}, 2(1):197, 1960.

\bibitem{DietrichFrisch:1990}
W.~Dieterich, H.~L. Frisch, and A.~Majhofer.
\newblock Nonlinear diffusion and density functional theory.
\newblock {\em Phys. Rev. B}, 78(2):317, 1990.

\bibitem{Ebner:1977sw}
C.~Ebner and W.~F. Saam.
\newblock New phase-transition phenomena in thin {A}rgon films.
\newblock {\em Phys. Rev. Lett.}, 38(25):1486, 1977.

\bibitem{elGendi:1969}
S.~E. El-Gendi.
\newblock Chebyshev solution of differential, integral and integro-differential
  equations.
\newblock {\em Comput. J.}, 12(3):282, 1969.

\bibitem{Evans}
R.~Evans.
\newblock The nature of the liquid-vapour interface and other topics in the
  statistical mechanics of non-uniform, classical fluids.
\newblock {\em Adv. Phys.}, 28(2):143, 1979.

\bibitem{RosenfeldObituary:2002}
R.~Evans, J.-P. Hansen, and H.~L\"{o}wen.
\newblock Foreword.
\newblock {\em J. Phys.: Condens. Matter}, 14(46), 2002.
\newblock \url{http://stacks.iop.org/0953-8984/14/i=46/a=001}.

\bibitem{FrinkSalinger:2000kx}
L.~J.~D. Frink and A.~G. Salinger.
\newblock Two- and three-dimensional nonlocal density functional theory for
  inhomogeneous fluids: {I}. {A}lgorithms and parallelization.
\newblock {\em J. Comput. Phys.}, 159(2):407, 2000.

\bibitem{DouglasFrink:2000425}
L.~J.~D. Frink and A.~G. Salinger.
\newblock Two- and three-dimensional nonlocal density functional theory for
  inhomogeneous fluids: {II}. {S}olvated polymers as a benchmark problem.
\newblock {\em J. Comput. Phys.}, 159(2):425, 2000.

\bibitem{FrinkSalinger:2002:NumericalChallenges}
L.~J.~D. Frink, A.~G. Salinger, M.~P. Sears, J.~D. Weinhold, and A.~L.
  Frischknecht.
\newblock Numerical challenges in the application of density functional theory
  to biology and nanotechnology.
\newblock {\em J. Phys.: Condens. Matter}, 14(46):12167, 2002.

\bibitem{Goddard:2013multi}
B.~D. Goddard, A.~Nold, and S.~Kalliadasis.
\newblock Multi-species dynamical density functional theory.
\newblock {\em J. Chem. Phys.}, 138(14):144904, 2013.

\bibitem{Goddard:2012general}
B.~D. Goddard, A.~Nold, N.~Savva, G.~A. Pavliotis, and S.~Kalliadasis.
\newblock General dynamical density functional theory for classical fluids.
\newblock {\em Phys. Rev. Lett.}, 109(12):120603, 2012.

\bibitem{Goddard:2013Unification}
B.~D. Goddard, A.~Nold, N.~Savva, P.~Yatsyshin, and S.~Kalliadasis.
\newblock Unification of dynamic density functional theory for colloidal fluids
  to include inertia and hydrodynamic interactions: derivation and numerical
  experiments.
\newblock {\em J. Phys.: Condens. Matter}, 25(3):035101, 2013.

\bibitem{BenDDFT}
B.~D. Goddard, G.~A. Pavliotis, and S.~Kalliadasis.
\newblock The overdamped limit of dynamic density functional theory: {R}igorous
  results.
\newblock {\em SIAM Multiscale Model. Simul.}, 10(2):633--663, 2012.

\bibitem{Gor:2012:QuenchedSolid}
G.~Y. Gor, M.~Thommes, K.~A. Cychosz, and A.~V. Neimark.
\newblock Quenched solid density functional theory method for characterization
  of mesoporous carbons by nitrogen adsorption.
\newblock {\em Carbon}, 50(4):1583, 2012.

\bibitem{Groot:1987uq}
R.~D. Groot, N.~M. Faber, and J.~P. v.~d. Eerden.
\newblock Hard sphere fluids near a hard wall and a hard cylinder.
\newblock {\em Mol. Phys.}, 62(4):861, 1987.

\bibitem{Grosch:1977273}
C.~E. Grosch and S.~A. Orszag.
\newblock Numerical solution of problems in unbounded regions: Coordinate
  transforms.
\newblock {\em J. Comput. Phys.}, 25(3):273, 1977.

\bibitem{Gross:2009:DFTSAFT}
J.~Gross.
\newblock {A density functional theory for vapor-liquid interfaces using the
  PCP-SAFT equation of state}.
\newblock {\em J. Chem. Phys.}, 131(20):204705, 2009.

\bibitem{GuoLabrosse:2013}
W.~Guo, G.~Labrosse, and R.~Narayanan.
\newblock {\em The application of the {C}hebyshev-spectral method in transport
  phenomena}.
\newblock Springer Science \& Business Media, 2013.

\bibitem{Haertel:2015:DFTDoublelayer}
A.~H\"{a}rtel, M.~Janssen, S.~Samin, and R.~v. Roij.
\newblock Fundamental measure theory for the electric double layer:
  implications for blue-energy harvesting and water desalination.
\newblock {\em J. Phys.: Condens. Matter}, 27(19):194129, 2015.

\bibitem{heroux:2007vn}
M.~A. Heroux, A.~G. Salinger, and L.~J.~D. Frink.
\newblock Parallel segregated {S}chur complement methods for fluid density
  functional theories.
\newblock {\em SIAM J. Sci. Comput.}, 29(5):2059, 2007.

\bibitem{Herring:2010vn}
A.~R. Herring and J.~R. Henderson.
\newblock Simulation study of the disjoining pressure profile through a
  three-phase contact line.
\newblock {\em J. Chem. Phys.}, 132(8):084702, 2010.

\bibitem{Hughes:2013:Water}
J.~Hughes, E.~J. Krebs, and D.~Roundy.
\newblock A classical density-functional theory for describing water
  interfaces.
\newblock {\em J. Chem. Phys.}, 138(2):024509, 2013.

\bibitem{Knepley:2010}
M.~G. Knepley, D.~A. Karpeev, S.~Davidovits, R.~S. Eisenberg, and D.~Gillespie.
\newblock An efficient algorithm for classical density functional theory in
  three dimensions: Ionic solutions.
\newblock {\em J. Chem. Phys.}, 132(12):124101, 2010.

\bibitem{konig:2005curvature}
P.-M. K{\"o}nig, P.~Bryk, K.~Mecke, and R.~Roth.
\newblock Curvature expansion of density profiles.
\newblock {\em Europhys. Lett.}, 69(5):832, 2005.

\bibitem{Lang:2001:thesis}
A.~Lang.
\newblock {Doctoral Thesis Technical University of Vienna}, 2001.

\bibitem{Le:1992}
P.~Le~Qu{\'e}r{\'e}, R.~Masson, and P.~Perrot.
\newblock {A Chebyshev collocation algorithm for 2D non-Boussinesq convection}.
\newblock {\em J. Comput. Phys.}, 103(2):320, 1992.

\bibitem{Mai:2007}
N.~Mai-Duy and R.~I. Tanner.
\newblock A spectral collocation method based on integrated {C}hebyshev
  polynomials for two-dimensional biharmonic boundary-value problems.
\newblock {\em J. Comput. Appl. Math.}, 201(1):30, 2007.

\bibitem{Malijevsky:2013:CriticalPointWedgeFilling}
A.~Malijevsk\'y and A.~O. Parry.
\newblock Critical point wedge filling.
\newblock {\em Phys. Rev. Lett.}, 110(16):166101, 2013.

\bibitem{Marconi:2007zr}
U.~M.~B. Marconi and S.~Melchionna.
\newblock Phase-space approach to dynamical density functional theory.
\newblock {\em J. Chem. Phys.}, 126(18):184109, 2007.

\bibitem{Marconi:1999ys}
U.~M.~B. Marconi and P.~Tarazona.
\newblock Dynamic density functional theory of fluids.
\newblock {\em J. Chem. Phys.}, 110(16):8032, 1999.

\bibitem{Marconi:2000ys}
U.~M.~B. Marconi and P.~Tarazona.
\newblock Dynamic density functional theory of fluids.
\newblock {\em J. Phys.: Condens. Matter}, 12(8A):A413, 2000.

\bibitem{Marconi:2006zr}
U.~M.~B. Marconi and P.~Tarazona.
\newblock Nonequilibrium inertial dynamics of colloidal systems.
\newblock {\em J. Chem. Phys.}, 124(16):164901, 2006.

\bibitem{MaroniTarazona:2008:Beyond}
U.~M.~B. Marconi, P.~Tarazona, F.~Cecconi, and S.~Melchionna.
\newblock Beyond dynamic density functional theory: the role of inertia.
\newblock {\em J. Phys.: Condens. Matter}, 20(49):494233, 2008.

\bibitem{DissMerath:2008}
R.-J.~C. Merath.
\newblock {\em Microscopic calculations of line tensions}.
\newblock PhD thesis, Max-Planck-Institut f{\"u}r Metallforschung Stuttgart,
  2008.

\bibitem{Mermin:1965fk}
N.~D. Mermin.
\newblock Thermal properties of the inhomogeneous electron gas.
\newblock {\em Phys. Rev.}, 137(5A), 03 1965.

\bibitem{Mitchell2001}
P.~Mitchell et~al.
\newblock Microfluidics-downsizing large-scale biology.
\newblock {\em Nature biotechnology}, 19(8):717--721, 2001.

\bibitem{Neal03}
R.~M. Neal.
\newblock Slice sampling.
\newblock {\em Ann. Stat.}, 31(3):705, 2003.

\bibitem{Nold:2016:thesis}
A.~Nold.
\newblock {\em {PhD thesis: From the nano- to the macroscale - Bridging scales
  for the moving contact line problem}}.
\newblock Imperial College London, 2016.

\bibitem{Nold:FluidStructure:2014}
A.~Nold, D.~N. Sibley, B.~D. Goddard, and S.~Kalliadasis.
\newblock Fluid structure in the immediate vicinity of an equilibrium
  three-phase contact line and assessment of disjoining pressure models using
  density functional theory.
\newblock {\em Phys. Fluids}, 26(7):072001, 2014.

\bibitem{nold2015nanoscale}
A.~Nold, D.~N. Sibley, B.~D. Goddard, and S.~Kalliadasis.
\newblock Nanoscale fluid structure of liquid-solid-vapour contact lines for a
  wide range of contact angles.
\newblock {\em Math. Model. Nat. Phenom.}, 10(4):111--125, 2015.

\bibitem{Parry:2014:WetGroove}
A.~O. Parry, A.~Malijevsk\'y, and C.~Rasc\'on.
\newblock Capillary contact angle in a completely wet groove.
\newblock {\em Phys. Rev. Lett.}, 113:146101, 2014.

\bibitem{Antonio2010}
A.~Pereira and S.~Kalliadasis.
\newblock Equilibrium gas-liquid-solid contact angle from density-functional
  theory.
\newblock {\em J. Fluid Mech.}, 692:53--77, 2012.

\bibitem{Rex:2009qf}
M.~Rex and H.~L{\"o}wen.
\newblock Dynamical density functional theory for colloidal dispersions
  including hydrodynamic interactions.
\newblock {\em Eur. Phys. J. E}, 28(2):139, 2009.

\bibitem{Rosenfeld:1989qc}
Y.~Rosenfeld.
\newblock Free-energy model for the inhomogeneous hard-sphere fluid mixture and
  density-functional theory of freezing.
\newblock {\em Phys. Rev. Lett.}, 63(9):980, 1989.

\bibitem{rosenfeld1996dimensional}
Y.~Rosenfeld, M.~Schmidt, H.~L{\"o}wen, and P.~Tarazona.
\newblock Dimensional crossover and the freezing transition in density
  functional theory.
\newblock {\em J. Phys.: Condens. Matter}, 8(40):L577, 1996.

\bibitem{RosenfeldTarazona:1997}
Y.~Rosenfeld, M.~Schmidt, H.~L\"owen, and P.~Tarazona.
\newblock Fundamental-measure free-energy density functional for hard spheres:
  Dimensional crossover and freezing.
\newblock {\em Phys. Rev. E}, 55(4):4245, 1997.

\bibitem{Roth:2010fk}
R.~Roth.
\newblock Fundamental measure theory for hard-sphere mixtures: a review.
\newblock {\em J. Phys.: Condens. Matter}, 22(6):063102, 2010.

\bibitem{RothEvansLang:2002}
R.~Roth, R.~Evans, A.~Lang, and G.~Kahl.
\newblock Fundamental measure theory for hard-sphere mixtures revisited: the
  white bear version.
\newblock {\em J. Phys.: Condens. Matter}, 14(46):12063, 2002.

\bibitem{RothMeckeOettel2012}
R.~Roth, K.~Mecke, and M.~Oettel.
\newblock Communication: Fundamental measure theory for hard disks: Fluid and
  solid.
\newblock {\em J. Chem. Phys.}, 136(8):081101, 2012.

\bibitem{Salzer:1972}
H.~E. Salzer.
\newblock Lagrangian interpolation at the {C}hebyshev points xn, $\nu$$\equiv$
  cos ($\nu$$\pi$/n), $\nu$= 0 (1) n; some unnoted advantages.
\newblock {\em Comput. J.}, 15(2):156, 1972.

\bibitem{Sears:2003uq}
M.~P. Sears and L.~J.~D. Frink.
\newblock A new efficient method for density functional theory calculations of
  inhomogeneous fluids.
\newblock {\em J. Comput. Phys.}, 190(1):184, 2003.

\bibitem{Shen2009}
J.~Shen and L.~Wang.
\newblock Some recent advances on spectral methods for unbounded domains.
\newblock {\em Commun. Comput. Phys.}, 5(2):195, 2009.

\bibitem{TangWu:2003}
Y.~Tang and J.~Wu.
\newblock {A density-functional theory for bulk and inhomogeneous
  Lennard--Jones fluids from the energy route}.
\newblock {\em J. Chem. Phys.}, 119(14):7388, 2003.

\bibitem{Tarazona:2000:Freezing}
P.~Tarazona.
\newblock Density functional for hard sphere crystals: A fundamental measure
  approach.
\newblock {\em Phys. Rev. Lett.}, 84(4):694, 2000.

\bibitem{tarazona2008beyond}
P.~Tarazona and U.~M.~B. Marconi.
\newblock Beyond the dynamic density functional theory for steady currents:
  Application to driven colloidal particles in a channel.
\newblock {\em J. Chem. Phys.}, 128(16):164704, 2008.

\bibitem{Tee:2006Ad}
T.~W. Tee.
\newblock {\em An adaptive rational spectral method for differential equations
  with rapidly varying solutions}.
\newblock PhD thesis, University of Oxford, 2006.

\bibitem{Trefethen_2000}
N.~L. Trefethen.
\newblock {\em Spectral Methods in {MATLAB}}.
\newblock SIAM, Philadelphia, 2000.

\bibitem{SwolHenderson:1989}
F.~van Swol and J.~R. Henderson.
\newblock Wetting and drying transitions at a fluid-wall interface:
  Density-functional theory versus computer simulation.
\newblock {\em Phys. Rev. A}, 40(5):2567, 1989.

\bibitem{Wu-DFT}
J.~Wu.
\newblock Density functional theory for chemical engineering: {F}rom
  capillarity to soft materials.
\newblock {\em AIChE J.}, 52(3):1169, 2006.

\bibitem{Wu:2007fh}
J.~Wu and Z.~Li.
\newblock Density-functional theory for complex fluids.
\newblock {\em Annu. Rev. Phys. Chem.}, 58(1):85, 2007.

\bibitem{YatsyshinSerafim:2012}
P.~Yatsyshin, N.~Savva, and S.~Kalliadasis.
\newblock Spectral methods for the equations of classical density-functional
  theory: {R}elaxation dynamics of microscopic films.
\newblock {\em J. Chem. Phys.}, 136(12):124113, 2012.

\bibitem{PeterPRE}
P.~Yatsyshin, N.~Savva, and S.~Kalliadasis.
\newblock Geometry-induced phase transition in fluids: {C}apillary prewetting.
\newblock {\em Phys. Rev. E}, 87:020402(R), 2013.

\bibitem{PeterJPhysCondMatt}
P.~Yatsyshin, N.~Savva, and S.~Kalliadasis.
\newblock Density functional study of condensation in capped capillaries.
\newblock {\em J. Phys.: Condens. Matter}, 27(27):275104, 2015.

\bibitem{PeterJChemPhys2}
P.~Yatsyshin, N.~Savva, and S.~Kalliadasis.
\newblock Wetting of prototypical one- and two-dimensional systems:
  {T}hermodynamics and density functional theory.
\newblock {\em J. Chem. Phys.}, 142(3):034708, 2015.

\bibitem{YuWu:2002}
Y.-X. Yu and J.~Wu.
\newblock Density functional theory for inhomogeneous mixtures of polymeric
  fluids.
\newblock {\em J. Chem. Phys.}, 117(5):2368, 2002.

\bibitem{YuWu:2002:Modified}
Y.-X. Yu and J.~Wu.
\newblock Structures of hard-sphere fluids from a modified fundamental-measure
  theory.
\newblock {\em J. Chem. Phys.}, 117(22):10156, 2002.

\bibitem{Zhou:2012three}
D.~Zhou, J.~Mi, and C.~Zhong.
\newblock Three-dimensional density functional study of heterogeneous
  nucleation of droplets on solid surfaces.
\newblock {\em J. Phys. Chem. B}, 116(48):14100, 2012.

\bibitem{Zwanzig:1954cq}
R.~W. Zwanzig.
\newblock High-temperature equation of state by a perturbation method. {I.}
  {N}onpolar gases.
\newblock {\em J. Chem. Phys.}, 22(8):1420, 1954.

\end{thebibliography}

\appendix
\section{Convolution weights for 2D geometries \label{sec:ConvWeights2D}}

For a system of hard spheres, the DFT formulation includes at several
instances convolutions in 3D: In the definition of the weighted densities
$\nDensity_\alpha$ in \refe{eq:nAlphaDef}, as well as in the hard-sphere and
the attractive contributions to the free energy
\refe{eq:Rosenfeld_Variational} and \refe{eq:AttrContrEulerLagrange},
respectively. If the system is invariant in one or more directions, these 3D
convolutions may be rewritten as lower-dimensional convolutions. In
particular, if the system is invariant in one direction in Cartesian
coordinates, then a convolution $\klamm{\nDensity \ast \phi}_{\text{3D}}$ can
be rewritten as a 2D convolution $\klamm{\nDensity \ast
\PhiTwoD}_{\text{2D}}$ in the $y_1$-$y_2$-plane with
\begin{align}
\PhiTwoD\klamm{y_1,y_2} &= \int_{-\infty}^\infty \phi\klamm{y_1,y_2,y_3} \dI y_3. \label{eq:2Dweight}
\end{align}
The corresponding 2D-weight function for the attractive contribution to the free energy
as defined in \refe{eq:pattr} is for $r<1$
\begin{align}
\PhiTwoD^{\text{attr}}(r) &= \frac{\depthLJ}{\eD}\klammCurl{
\frac{3 \sqrt{1-r^2}}{160 r^{10}} \klamm{-105 -70r^2 -56r^4 +112 r^6 + 64 r^8} - \frac{3 \arcsin (r)}{32 r^{11}} \klamm{32 r^6 - 21}}, \label{eq:phiattr:2D:1}
\end{align}
and for $1 \leq r < r_c$, with cutoff-length $r_c$
\begin{align}
\PhiTwoD^{\text{attr}}(r) &= \frac{\depthLJ}{\eD}\pi \klamm{\frac{63}{64}\frac{1}{r^{11}} - \frac{3}{2}\frac{1}{r^5}}. \label{eq:phiattr:2D:2}
\end{align}
Beyond the cutoff length, $r \geq r_c$, we set $\PhiTwoD^{\text{attr}}(r)=0$.
We note that for small values of $r$, we evaluate a Taylor expansion of $\Phi_{2D}^{\text{attr}}$ at zero for computational purposes.
In dimensionless form, \refe{eq:defineEnergyScale} becomes
\begin{align}
\int_{\mathbb{R}^3} \BHattr\klamm{|\pos|} \dI \pos = - \frac{32}{9} \pi,
\end{align}
hence linking the dimensionless quantity $\frac{\eD}{\LJdepth}$ to the dimensionless cutoff-radius $r_c$ through
\begin{align}
\frac{\eD}{\LJdepth} = 1 - \frac{9}{32}\pi \klamm{ \frac{1}{r_c^3} - \frac{7}{32 r_c^9}}.
\end{align}
Finally, the convolution with FMT weight $\weight_{3,i}$ defined in
\refe{eq:WeightedDensitiesRFMT2} can be rewritten in 2D as a convolution with
\begin{align}
\PhiTwoD^{\weight_{3,i}}(r) = 2 \sqrt{R_i^2 - r^2} \Theta\klamm{R_i-r}, \label{eq:phi2D:w3i:def}
\end{align}
where $r = \sqrt{y_1^2 + y_2^2}$.

\section{Implementation of pseudospectral methods in 2D \label{sec:NumericalRepresentation}}
The collocation points ${\bf x}$ and the function $f({\bf x})$ in the 2D
computational domain are represented in a block-structure as
\begin{align}
{\bf x} =
\begin{pmatrix}
(x_1^1,x_2^1)\\
(x_1^1,x_2^2)\\
\ldots\\
(x_1^1,x_2^{N_2})\\\hline
(x_1^2,x_2^1)\\
\ldots\\
(x_1^2,x_2^{N_2})\\\hline
\ldots\\
\ldots\\
(x_1^{N_1},x_2^{N_2})\\
\end{pmatrix}
\qquad \text{and}\qquad
{\bf f} = 
 \begin{pmatrix}
f(x_1^1,x_2^1)\\
f(x_1^1,x_2^2)\\
\ldots\\
f(x_1^1,x_2^{N_2})\\\hline
f(x_1^2,x_2^1)\\
\ldots\\
f(x_1^2,x_2^{N_2})\\\hline
\ldots\\
\ldots\\
f(x_1^{N_1},x_2^{N_2})\\
\end{pmatrix}, \label{eq:Spectral:FunctionVector:f}
\end{align}
where $x_j^i$ denotes the $i$th collocation point of coordinate $j$. Any
linear operator ${\bf A} = \klamm{a}_{ij}$ acting on $x_1$ can be written in
2D
as a tensor product
\begin{align}
{\bf A} \otimes {\bf I}_{N_2}=
\begin{pmatrix}
a_{11} & a_{12}& \ldots\\ a_{21}& \ldots\\ \dots&& a_{N_1,N_1}
\end{pmatrix}
\otimes{\bf I}_{N_2}&=
\begin{pmatrix}
a_{11}{\bf I}_{N_2} & a_{12} {\bf I}_{N_2} & \ldots\\
 a_{21} {\bf I}_{N_2} & \ddots\\
 \ldots&0& a_{N_1,N_2} {\bf I}_{N_2}
\end{pmatrix},
\end{align}
which can then be applied on the function vector
(\ref{eq:Spectral:FunctionVector:f}). Similarly, an operator acting on $x_2$
can be written as
\begin{align}
{\bf I}_{N_1} \otimes {\bf B} =
{\bf I}_{N_1} \otimes
\begin{pmatrix}
b_{11} & b_{12}& \ldots\\ b_{21}& \ldots\\ \dots&& b_{N_2,N_2}
\end{pmatrix} =
\begin{pmatrix}
{\bf B} & 0& \\
0 & \ddots & \\
0 & \ldots & {\bf B}
\end{pmatrix}.
\end{align}
This allows us to easily define linear interpolation, integration and
differentiation operators in 2D.

\subsection{Representation of the physical domain \label{sec:App:RepPhys}}

The physical domain is mapped onto the
Cartesian domain via $\CartMap$ given in Eq.~\refe{eq:DefCartMap}.
In this case, the integration vector
for the physical domain is computed through
\begin{align}
w_{\PhysSpace} = \klamm{w_{N_1} \otimes w_{N_2}} \text{diag}\klamm{\text{det}\klamm{\bf J}},
\end{align}
where $w_{N_{1,2}}$ are the 1D integration weights given in
\refe{eq:Num:Cheb:ClenshawCurtisQuad1}, and $\text{det}\klamm{\bf J} =
\diff{\PhysSpace_1}{x_1} \diff{\PhysSpace_2}{x_2} - \diff{\PhysSpace_1}{x_2}
\diff{\PhysSpace_2}{x_1}$ is the vector of the determinants of the Jacobian
at the collocation points ${\bf x}$.

We distinguish between the following three cases:
\begin{enumerate}
\item Skewed Cartesian coordinates, by angle $\alpha$. In this case,
\begin{align}
\CartMap_{\text{SC},\alpha}:
\begin{pmatrix} y_1'\\ y_2' \end{pmatrix}
\to
\begin{pmatrix}
1 & \cos \alpha\\
0 & \sin \alpha
\end{pmatrix}
\begin{pmatrix}
y_1'\\ y_2'
\end{pmatrix} \label{eq:SkewedCartesianCoordinates}
\end{align}
and the integration weight vector has to be rescaled by $\sin \alpha$:
\begin{align}
w_{SC} = w_\PhysSpace \sin \alpha.
\end{align}
This map has been applied when computing the density profile in the
vicinity of a contact
line~\cite{Nold:FluidStructure:2014,nold2015nanoscale}.
\item Polar coordinates:
\begin{align}
\CartMap_{\text{pol}} :
\begin{pmatrix} \varphi\\ r \end{pmatrix}
\to
\begin{pmatrix} r \cos \varphi \\ r \sin \varphi \end{pmatrix}.
\end{align}
$\varphi \in [0,2\pi)$ and $r \in [0,\infty]$ are the angular and the radial variables, respectively. The integration vector has to be multiplied with $r$:
\begin{align}
w_{\text{pol}} = w_\PhysSpace \text{diag}\klamm{r}.
\end{align}
\item Projected spherical coordinates with radial parameter $R$:
\begin{align}
\CartMap_{\text{sph}} :
\begin{pmatrix} \theta \\ \psi \end{pmatrix}
\to
\begin{pmatrix}
R \cos \psi \sin \theta \\ R \cos \theta
\end{pmatrix},
\end{align}
where $\theta \in [0,\pi]$ and $\psi \in [0,2\pi]$ are angular
variables. We distinguish two cases: First, for an integration over the 2D
disk defined by $\CartMap_{\text{sph}}$, the integration weight vector has
to be multiplied with the determinant of the Jacobian
\begin{align}
w_{\text{sph}} = w_\PhysSpace R^2\text{diag}\klamm{\sin^2\klamm{\theta} \sin\klamm{\psi}}.
\end{align}
If, however, the map $\CartMap_{\text{sph}}$ is complemented by a third cartesian component $y_3 = R \sin \psi \sin \theta$, and the integration is
done over the surface of the sphere with radius $R$ whilst assuming invariance of the density profile in $y_3$-direction, then the integration weight vector is multiplied with the classical weight for spherical coordinates
\begin{align}
w_{\text{surf}} = w_\PhysSpace R^2\text{diag}\klamm{\sin\klamm{\theta}}.
\end{align}
\end{enumerate}

\subsection{Domain discretisation  \label{Discretization}}
For the computation of the convolutions introduced in Sec.~\ref{S:convolution}, the intersections between the supports of the weight functions and of the density profile have to be discretised. Here, we present the maps from the computational domain to the physical domain needed for the discretisation of the different intersections.

First, we are interested in the density distribution of a fluid in contact with a planar hard wall. The domain which needs to be discretised is the half-space $\HalfSpace$. 
We therefore choose to map the domain $[-1,1]^2$, discretised by Gauss-Lobatto-Chebyshev collocation points in 2D, onto the half-space, using Eq.~(\ref{eq:General2DMapFromCompDomain}) and employing the algebraic maps defined in Eqs.~(\ref{eq:WholeLineAlgebraicMapping}) and (\ref{eq:SemiInfiniteAlgMapping}) for the first and second variables $y_1'$ and $y_2'$, respectively:
\begin{align}
\PhysSpace_{\HalfSpace}(x_1,x_2) \defi
\klamm{\AlgMapWhole(x_1),\AlgMapSemi(x_2)}. \label{eq:HalfSpaceDiscretization}
\end{align} 

When computing the convolutions in real-space, the intersection $\Intersection(y)$ between the support of the weight function $\convWeight$ and the shifted half-space needs to be discretised separately. 
For the weights $\PhiTwoDAttr$, $\PhiTwoD^{\weight_{3,i}}$ and $\weight_{2,i}$, given in Eqs. (\ref{eq:phiattr:2D:1})-(\ref{eq:phiattr:2D:2}), (\ref{eq:phi2D:w3i:def}), and (\ref{eq:WeightedDensitiesRFMT2}), respectively, we distinguish four cases: 
\begin{enumerate}[(1)]
\item The support of the weight function corresponds to the surface of a sphere of radius $R$ in 3D: $A_1 \defi \{\pos \in \mathbb{R}^3: |\pos| = R \}$, and the density under consideration is invariant in 1D. This is a special case of (2), and is applicable for weights $\weight_{2,i}$ and ${\boldsymbol \weight}_{2,i}$ given in Eqs.~(\ref{eq:WeightedDensitiesRFMT2}).
\item The support of the weight function corresponds to a disk of radius $R$ in 2D: $A_2 \defi \{\pos \in \mathbb{R}^2: |\pos| < R \}$. This is used for weight $\weight_{3,i}$ in Eq.~(\ref{eq:WeightedDensitiesRFMT2}) if the density is invariant in one direction, and for the inner contribution to the 2D projection of the attractive particle--particle interaction (see Eqs.~(\ref{eq:phiattr:2D:1})).
\item The support of the weight function corresponds to an infinite annulus with inner radius $R$: $A_3 \defi \{\pos \in \mathbb{R}^2: |\pos| > R \}$. This is applicable for the outer contribution to the the 2D projection of the attractive particle--particle interaction without radial cutoff, $\PhiTwoDAttr$ (see Eqs.~(\ref{eq:phiattr:2D:2})).
\item The support of the weight function corresponds to a finite annulus with inner radius $R_{\text{in}}$ and outer radius $R_{\text{out}}$: $A_4 \defi \{\pos \in \mathbb{R}^2: R_{\text{in}} \leq |\pos| \leq R_{\text{out}}  \}$.  This is applicable for the outer contribution to the the 2D projection of the attractive particle--particle interaction with a finite cutoff radius $r_c$ (see Eqs.~(\ref{eq:phiattr:2D:2})).
\end{enumerate}
In Tabs.~\ref{tab:FourierShapes} - \ref{tab:ShapeDiscretization}, we present discretisations of eleven different geometries. 
In Table~\ref{tab:AssemblingShapesForIntersections}, it is then described how these geometries are assembled to represent the intersections $\Intersection\klamm{\pos}$ between $A_{1-4}$ with the shifted half-space.

For the discretisation of a disk, a finite annulus or an infinite annulus (see Table~\ref{tab:FourierShapes}), we take advantage of the periodicity in the angular variable in the polar coordinates and employ an equispaced grid to describe the second computational variable $\hat x_2 \in [0,1)$, which is then mapped to the angular variable $\varphi \in [0,2\pi)$. For the discretisation of a disk with radius $R$, $\Disc\klamm{R}$, the radial variable is described by an even number of Gauss-Lobatto-Chebyshev collocation points, and mapped from $[-1,1] \to [-R,R]$. 
Analogous to the procedure employed in Ref.~\cite{Trefethen_2000}, this avoids the treatment of the origin, but requires the explicit implementation of the symmetry $f(r,\varphi) = f(-r,\text{mod}(\varphi+\pi,2\pi))$, hence leading to an effective restriction of the computational variable $x_1$ to $[0,1]$, and restoring bijectivity of the map $\PhysSpace_{\Disc\klamm{R}}$. Alternatively, a Radau grid may be employed, which includes only one boundary point, therefore also avoiding the explicit treatment of the origin~\cite{Boyd:2001Cheb,GuoLabrosse:2013}.

The geometries described in Table~\ref{tab:ShapeDiscretization:Spherical} map the computational domain 
onto spherical angular coordinates. The discretisation of $\Ball$ presented there uses the periodicity in the second variable, which is useful for the computation of the weighted density $\nDensity_2$ (see Eq.~\refe{eq:nAlphaDef}).
In particular, $\Ball\klamm{R,\theta_1,\theta_2}$ is the surface of a sphere with radius $R$, capped at $\theta_1$ 
and $\theta_2$, respectively. If the full surface of the sphere is to be described, then $\theta_1 = 0$ and $\theta_2 = \pi$. 
The convolution of the density distribution with the second FMT weight $\weight_2$ is then written as
\begin{align}
\int_{\mathbb{R}^3} \delta\klamm{R-|\tilde \pos|} \nDensity_\pos\klamm{\tilde \pos} \dI \tilde \pos
&= \int_{\Ball\klamm{R,\theta_1,\theta_2}} \nDensity_\pos\klamm{\tilde \pos} \dI \tilde \pos \notag \\
&= R^2 \int_{\theta_1}^{\theta_2} \int_{0}^{2 \pi} \sin (\theta) \nDensity_\pos\klamm{ \CartMapSph\klamm{\theta,\psi} }  \dI \psi \dI \theta,
\label{eq:BallIntegration}
\end{align}
and similarly for the vector-weight ${\boldsymbol \weight}_2$. $\nDensity_\pos$ is defined in Eq.~\refe{eq:Numerics:DefineShiftedDensity}. Here, the 2D density profile $\nDensity$ was extended into the third dimension by assuming invariance in $y_{3}$. It is now clear that the integrand is periodic in $\psi$.

In contrast to $\Ball$, $\Sphere$ describes the 2D projection of $\Ball$ onto the $y_1$-$y_2$ plane. This is used to integrate 
the FMT weight $\PhiTwoD^{\weight_{3}}$ for a system invariant in one direction (see Eq.~\refe{eq:phi2D:w3i:def}). 
In particular, the integration is written in angular spherical coordinates as:
\begin{align}
\int_{\mathbb{R}^2} \PhiTwoD^{\weight_{3}}\klamm{\tilde \pos}\nDensity_\pos\klamm{\tilde \pos} \dI \tilde \pos
&= \int_{\Sphere\klamm{R,\theta_1,\theta_2}} 2 \sqrt{R^2- |\tilde \pos|^2} \nDensity_\pos\klamm{\tilde \pos} \dI \tilde \pos\notag\\
&= 2R^3 \int_{\theta_1}^{\theta_2} \int_{0}^{\pi} \sin^2(\theta) \sin^2 (\psi) \nDensity_\pos\klamm{ \CartMap_{\text{sph}}\klamm{\theta,\psi} }  \dI \psi \dI \theta,
\end{align}
where $R^2 \sin^2\klamm{\theta}\sin\klamm{\psi}$ is the determinant of the Jacobian of $\CartMap_{\text{sph}}$.

The remaining discretisations $\PhysSpace_{\mathcal{P}_{1-6}}$ described in Table~\ref{tab:ShapeDiscretization} employ a discretisation of the computational domain with Gauss-Lobatto-Chebyshev collocation points in both directions and map it onto polar coordinates. Crucially, the radial component for the maps $\PhysSpace_{\HalfStripMinusDisc}$ and $\PhysSpace_{\SlitMinusDisc}$ is mapped onto an interval, whose size depends on the angular variable $\varphi$ and which extends to infinity as $\varphi$ approaches the boundary. We therefore employ a generalisation of the algebraic map $\AlgMapSemi$ by defining 
\begin{subequations}
\begin{align}
\AlgMapSemiFinite(x;a,b,L) &=  a + (b-a) \frac{e}{2} \frac{1+x}{1-x+e},\\
\text{where} 
\qquad e &= 2 \frac{L}{(b-a)-2L}.
\end{align}
\label{eq:Numerics:DefQ}
\end{subequations}

Finally, we note that for discretisations $\PhysSpace_{\InfAnn}$ and $\PhysSpace_{\mathcal{P}_{1-3}}$, 
the parameter $L$ has to be estimated. For unbounded domains, $L$ determines the spreading rate of 
the collocation points. In particular, it defines the position of the red solid lines in Tabs.~\ref{tab:FourierShapes}-\ref{tab:ShapeDiscretization}.
Here, the density profile is assumed to change relatively slowly when compared to the weight function, and consequently $L$ is obtained by minimizing the maximum interpolation error of the weight function $\convWeight$ on the respective discretization.

\newcommand{\SketchWidth}{2.5cm}
\newcommand{\SketchWidthTable}{4cm}

\newpage
   \begin{longtable}{@{} lM{\SketchWidthTable} @{}} 
      \toprule
			Map from computational to physical domain& Discretisation in Cartesian coordinates \\\midrule			
$\PhysSpace_{\Disc\klamm{R}}:\begin{pmatrix} x_1\\\xE_2 \end{pmatrix} \to
\begin{pmatrix} r \\ \varphi \end{pmatrix} =
\begin{pmatrix}
R x_1\\
2\pi \xE_2 \textit{}
\end{pmatrix}$& 
\includegraphics{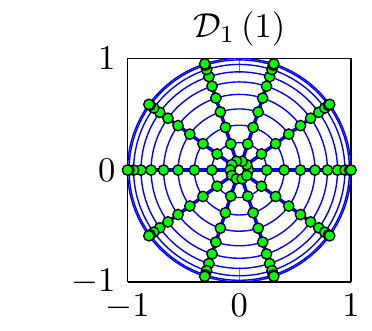}
\\\hline\\
$\PhysSpace_{\InfAnn\klamm{R,L}}: \begin{pmatrix} x_1\\\xE_2 \end{pmatrix} \to
\begin{pmatrix} r \\ \varphi \end{pmatrix} =
\begin{pmatrix}
R +  \AlgMapSemi(x_1;L)\\
2\pi \xE_2
\end{pmatrix}$ & 
\includegraphics{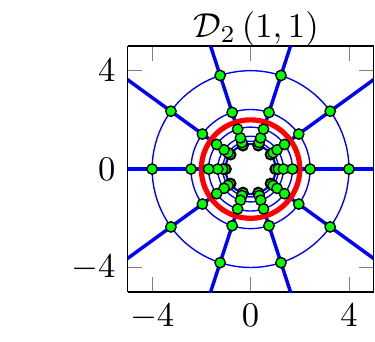}
\\\hline\\
$\PhysSpace_{\FinAnn\klamm{R_1,R_2}}: \begin{pmatrix} x_1\\\xE_2 \end{pmatrix} \to
\begin{pmatrix} r \\ \varphi \end{pmatrix} =
\begin{pmatrix}
R_1 +  \frac{x_1+1}{2}\klamm{R_2-R_1}\\
2\pi \xE_2
\end{pmatrix}$ & 
\includegraphics{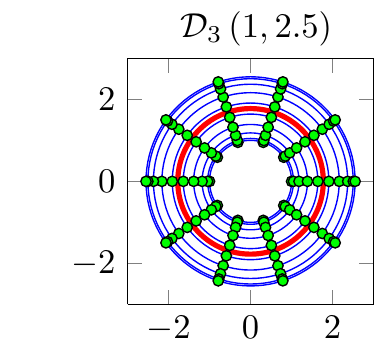}\\ 
\bottomrule
\caption{ \label{tab:FourierShapes}
Variable $x_1$ in the computational domain is discretized with Gauss-Lobatto-Chebyshev collocation points, $x_1 \in [-1,1]$,
while variable $\xE_2$ is discretized with an equispaced grid ($\xE_2 \in [0,1)$). Here, the physical domain represents polar coordinates $\klamm{r,\varphi}$.
If applicable, the red solid line denotes $\CartMap\klammCurl{\PhysSpace\klamm{0,\xE_2}}$, and each side of the red solid line contains $50\%$ of the collocation points.
}
\end{longtable}
\newpage
   \begin{longtable}{@{} lM{\SketchWidthTable} @{}}
         \toprule
	Map from computational to physical domain& Discretisation in Cartesian coordinates \\\midrule
 $ \PhysSpace_{\Ball\klamm{R, \theta_1,\theta_2}}:\begin{pmatrix} x_1\\ \xE_2 \end{pmatrix}\to
\begin{pmatrix} \theta \\ \psi \end{pmatrix} =
\begin{pmatrix}
\theta_1 + \klamm{\theta_2 - \theta_1} \frac{x_1+1}{2}
  \\ 2\pi \xE_2 \end{pmatrix} $ &  
\includegraphics{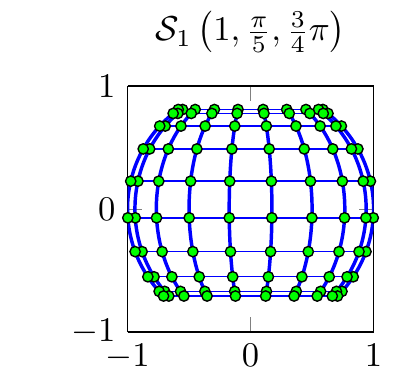}
  \\\hline\\
	$\PhysSpace_{\Sphere\klamm{R, \theta_1,\theta_2}}:\begin{pmatrix} x_1\\ x_2 \end{pmatrix}\to
\begin{pmatrix} \theta \\ \psi \end{pmatrix} =
\begin{pmatrix}
\theta_1 + \klamm{\theta_2 - \theta_1} \frac{x_2+1}{2}
  \\ \pi \frac{1+x_2}{2} \end{pmatrix} $ & 
\includegraphics{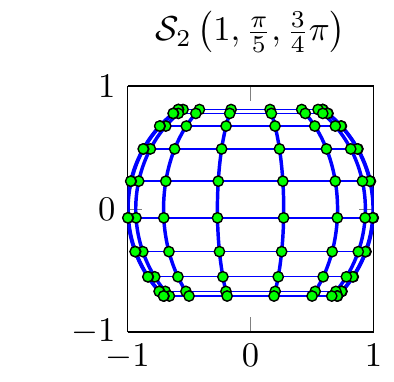}\\\bottomrule 
	\caption{ \label{tab:ShapeDiscretization:Spherical}
Variables $x_{1,2}$ in the computational domain are discretized with Gauss-Lobatto-Chebyshev collocation points, $x_{1,2} \in [-1,1]$,
    while variable $\xE_2$ is discretized with an equispaced grid ($\xE_2 \in [0,1)$). The physical domain represents angular spherical coordinates $\klamm{\theta,\psi}$.
}
\end{longtable}

   \begin{longtable}{@{}P{9.3cm}M{\SketchWidthTable} @{}} 
      \toprule
	Map from computational to physical domain
			& Discretisation in Cartesian coordinates \\\midrule			
 $\PhysSpace_{\AnnSeg\klamm{R,L}}:\begin{pmatrix} x_1\\x_2 \end{pmatrix} \to
\begin{pmatrix} r \\ \varphi \end{pmatrix} = 
\begin{pmatrix}
R + \AlgMapSemi(x_1;L)\\
\pi \frac{x_2+1}{2}
\end{pmatrix} $ & 
\includegraphics{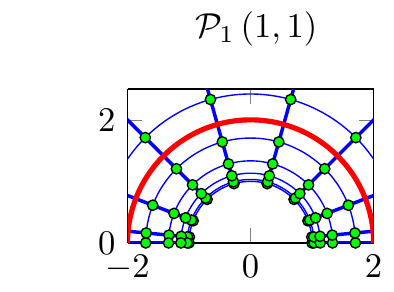}	\\ 
\\\hline
$\PhysSpace_{\HalfStripMinusDisc^\pm\klamm{R,L,h}}:
\begin{pmatrix} x_1\\x_2 \end{pmatrix} \to
\begin{pmatrix} r \\ \varphi \end{pmatrix}=
\begin{pmatrix}
\AlgMapSemiFinite\klamm{x_1;R,r_d,L_r}\\
\varphi_0^\pm + \klamm{\frac{\pi}{2}-\Delta \varphi} \frac{x_2+1}{2}
\end{pmatrix}$ &	
\includegraphics{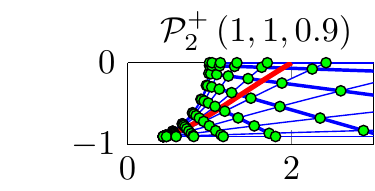}	\\
\\
{\small ($\varphi_0^- = \pi$, $\varphi_0^+ = \frac{3}{2}\pi+\Delta \varphi$, $\Delta \varphi = \arccos \left|\frac{h}{R} \right|$.)}\\\hline
$\PhysSpace_{\SlitMinusDisc\klamm{R,L,h}}:\begin{pmatrix} x_1\\x_2 \end{pmatrix} \to
\begin{pmatrix} r \\ \varphi \end{pmatrix}=
\begin{pmatrix}
\AlgMapSemiFinite\klamm{x_1;R,r_d,L_r}\\
\pi \frac{3+x_2}{2}
\end{pmatrix} $	&
\includegraphics{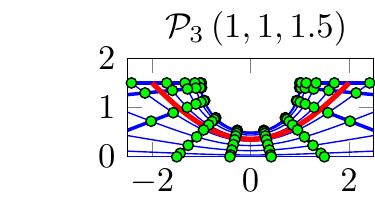}	\\
\\\hline\\
{\small $\PhysSpace_{\WedgeCutSide^\pm\klamm{R_{\text{in}},R_{\text{out}},h}}:
\begin{pmatrix} x_1\\x_2 \end{pmatrix} \to
\begin{pmatrix} r \\ \varphi \end{pmatrix} = 
\begin{pmatrix}
R_{\text{in}} + \frac{1+x_1}{2}\Delta r
\\
\frac{3}{2}\pi \pm
\varphi_0^\pm + \frac{1+x_2}{2} \klamm{\varphi^-_0 - \varphi^+_0}
\end{pmatrix}$}	&
\includegraphics{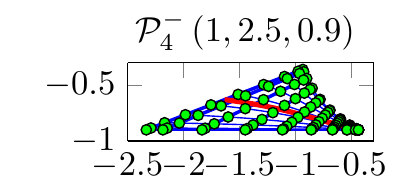}	\\  
{\small ($\varphi_0^- = \arccos \left|\frac{h}{R_{\text{out}}} \right|$, $\varphi_0^+ = \arccos \left|\frac{h}{R_{\text{in}}} \right|$, $ \Delta r  = \left|\frac{h}{\sin\varphi} \right| - R_{\text{in}} $.)}\\\hline
$\PhysSpace_{\WedgeCut\klamm{R_{\text{in}},R_{\text{out}},h}}:
\begin{pmatrix} x_1\\x_2 \end{pmatrix} \to
\begin{pmatrix} r \\ \varphi \end{pmatrix}= 
\begin{pmatrix}
R_{\text{in}} + \frac{1+x_1}{2}\Delta r 
\\
\frac{3}{2}\pi + \Delta \varphi
\end{pmatrix}$	&
\includegraphics{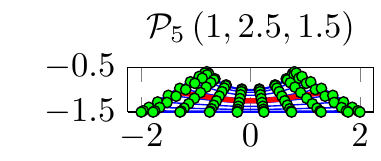}	\\ 
{\small ($\Delta \varphi = \arccos \frac{h}{R_{\text{out}}}$, 
$ \Delta r  = \left|\frac{h}{\sin\varphi} \right|-R_{\text{in}} $.)}
\\\hline
{\footnotesize $\PhysSpace_{\FiniteWedge\klamm{R_{\text{in}},R_{\text{out}},h}}:\begin{pmatrix} x_1\\x_2 \end{pmatrix} \to
\begin{pmatrix} r \\ \varphi \end{pmatrix}=
\begin{pmatrix}
R_{\text{in}} + \frac{1+x_1}{2}  \klamm{ R_{\text{out}} - R_{\text{in}}} \\
\frac{\pi}{2} + x_2 \klamm{ \pi - \Delta \varphi}
\end{pmatrix}$,\qquad($\Delta \varphi = \arccos\klamm{ \frac{h}{\Rout }}$.)}
	&
\includegraphics{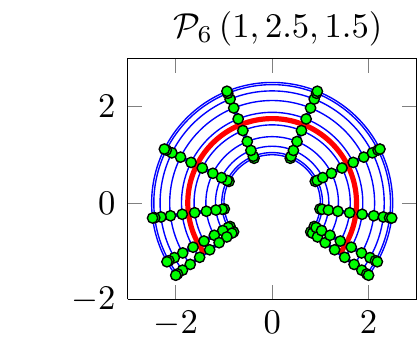}	\\
\bottomrule
\caption[Discretisations in polar coordinates, without periodic variable]{ \label{tab:ShapeDiscretization}
Variables $x_{1,2}$ in the computational domain are discretized with Gauss-Lobatto-Chebyshev collocation points, $x_{1,2} \in [-1,1]$. The physical domain represents polar coordinates $\klamm{r,\varphi}$.
We define $L_r  \defi L \frac{r_d-R}{3L+r_d-R}$ and $r_d \defi \left|\frac{h}{\sin \varphi}\right|$. Red line as in Table \ref{tab:FourierShapes}. $\AlgMapSemi$ and $\AlgMapSemiFinite$ are defined in Eqs.~\refe{eq:SemiInfiniteAlgMapping} and \refe{eq:Numerics:DefQ}, respectively.
}
\end{longtable}

\begin{table}[htbp]
   \centering
   \begin{tabular}{@{} lc @{}} 
      \toprule
			Support of weight $\convWeight$ & $\Intersection\klamm{y}$\\\midrule
		$\{{\bf r}: |{\bf r}|= R\}$ &$ 
  \left\{   \begin{array}{ll}
          \Ball\klamm{R,0,\pi} & \text{for } y> R\\
          \Ball\klamm{R,0,\frac{\pi}{2}+ \Delta \varphi}  & \text{for } y \in (-R,R)\\
 	 \emptyset  & \text{for } y < -R
     \end{array} \right\}
     $\\\hline		
		$\{{\bf r}: |{\bf r}| \leq R\}$ &$ 
  \left\{   \begin{array}{ll}
          \Sphere\klamm{R,0,\pi} & \text{for } y> R\\
          \Sphere\klamm{R,0,\frac{\pi}{2}+ \Delta \varphi}  & \text{for } y \in (-R,R)\\
 	 \emptyset  & \text{for } y < -R
     \end{array} \right\}
     $\\\hline		
			$\{{\bf r}: |{\bf r}| \geq R\}$ &
           $\left\{   \begin{array}{ll}
           \InfAnn\klamm{R,L}
       	& \text{for } y = \infty\\
	\AnnSeg\klamm{R,L} \cup
	\SlitMinusDisc\klamm{R,L,y}
	& \text{for } y \in [R,\infty)\\
		\AnnSeg\klamm{R,L} \cup 
	\klammCurl{\HalfStripMinusDisc^+\cup \HalfStripMinusDisc^-}\klamm{R,L,y}  
	& \text{for } y \in (0,R)\\
	\AnnSeg\klamm{R,L} & \text{for } y = 0
     \end{array} \right\}$ \\ \hline
			$\{{\bf r}: R_{\text{in}} \leq|{\bf r}| \leq R_{\text{out}}\}$ &
           $\left\{   \begin{array}{ll}
           \FinAnn\klamm{R_{\text{in}},R_{\text{out}}}
       	& \text{for } y \in [R_{\text{out}},\infty]\\			
				\klammCurl{\FiniteWedge \cup \WedgeCut}\klamm{R_{\text{in}},R_{\text{out}},y}  
	& \text{for } y \in (R_{\text{in}},R_{\text{out}})\\
	\klammCurl{\FiniteWedge \cup \WedgeCutSide^+ \cup \WedgeCutSide^-}\klamm{R_{\text{in}},R_{\text{out}},y} 
	& \text{for } y \in (0,R_{\text{in}}]\\	
	\FiniteWedge\klamm{R_{\text{in}},R_{\text{out}},y} & \text{for } y = 0
     \end{array} \right\}$ \\ 
           \bottomrule
   \end{tabular}
   \caption[Assembly of geometries defined in Tabs.~\ref{tab:FourierShapes}-\ref{tab:ShapeDiscretization}]{
	Assembly of geometries defined in Tabs.~\ref{tab:FourierShapes}-\ref{tab:ShapeDiscretization} to obtain the different
	intersections $\Intersection(y)$ of the support of weight $\convWeight$ with the shifted half-space $\{(x',y'):y + y' > 0\}$.	
	Here, $\Delta \varphi = \arcsin \frac{y}{R}$, and $L$ is a mapping parameter.}
   \label{tab:AssemblingShapesForIntersections}
\end{table}

\section{Multi species - convergence \label{sec:App:Multispecies}}

\begin{figure}[h]
	\includegraphics[width=\textwidth]{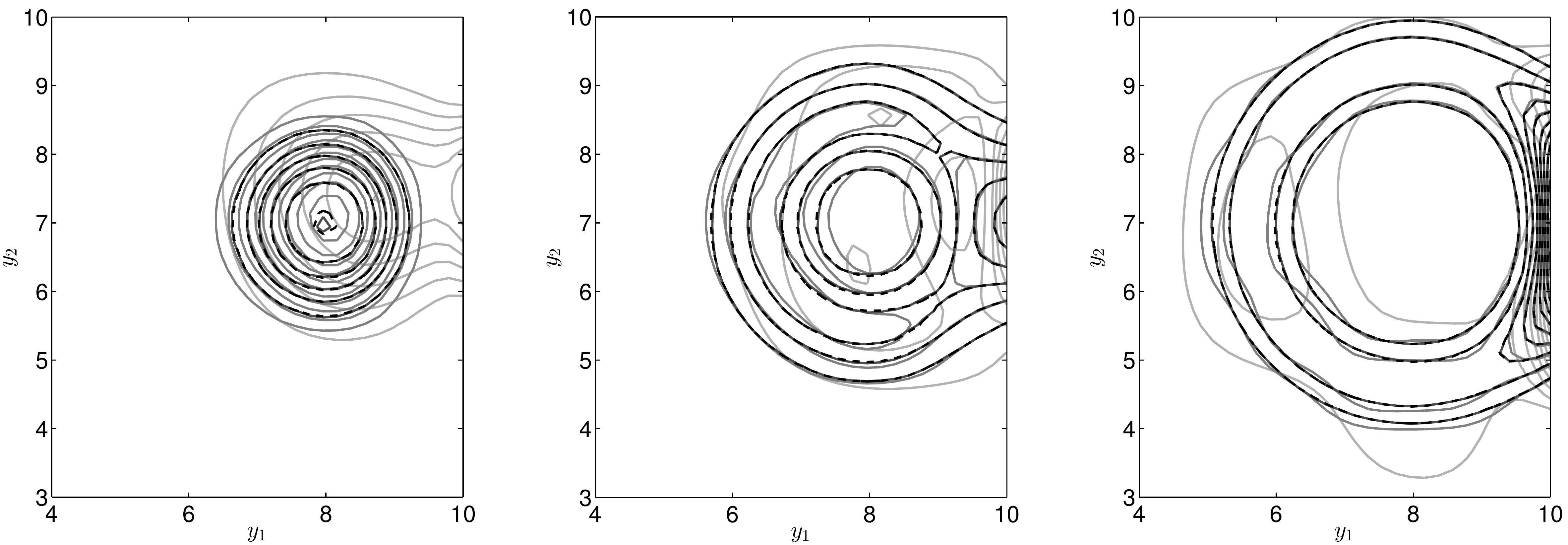}
	\caption{ \label{fig:App:multiSpeciesGaussianN} Density contours for the three Gaussian species also shown in Fig.~\ref{fig:multiSpeciesGaussian}, (increasing size
	left--right) for different numbers of collocation points $N=10,20,30,50$, represented by increasingly darker shades, where the curve representing $N=50$ (black) is also dashed. The curves for 30 and 50 collocation points
	are virtually indistinguishable. The particles are contained in a hard-wall box $[0,10] \times [0,10]$,
	but the plots are zoomed for clarity.}
\end{figure}

\begin{figure}[h]
	\includegraphics[width=\textwidth]{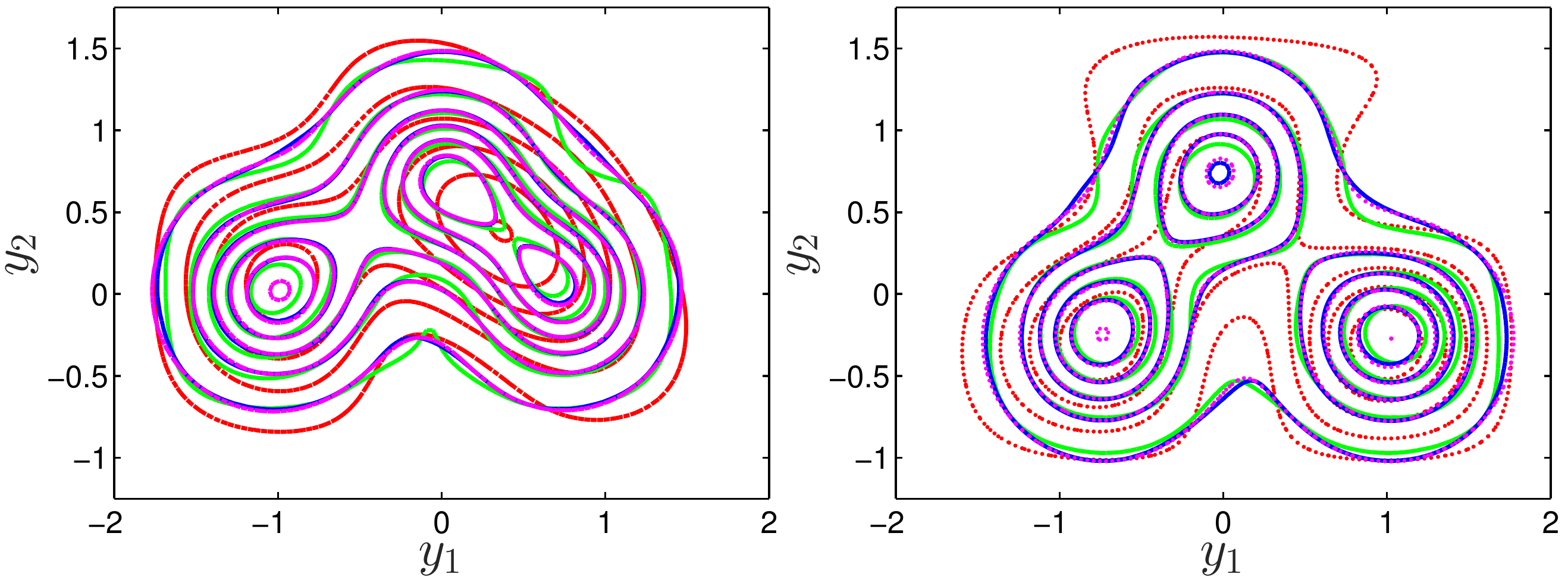}		
	\caption{ \label{fig:App:FMT2SpeciesN} Density contours for the two hard disks species also shown in Fig.~\ref{fig:FMT2Species}, subject to different external
	potentials, for different numbers of collocation points: $N=20$ (red, dotted), $N=30$, (green, solid),
	$N=40$ (blue, solid) and $N=60$ (magenta, dotted). The curves for 40 and 60 collocation points
	are virtually indistinguishable.}
\end{figure}

\end{document}